\pdfoutput=1
\documentclass[prd,twocolumn,showpacs,preprintnumbers,floatfix,nobibnotes,lengthcheck,balancelastpage]{revtex4}

\usepackage{graphicx}
\usepackage[raggedright,hang]{subfigure}
\usepackage{epstopdf}
\usepackage{rotating}
\usepackage{color}
\usepackage{hyperref}
\usepackage{amssymb}
\usepackage{amsmath}
\usepackage{amsfonts}
\usepackage{fancyhdr}
\usepackage{datetime}
\usepackage{times}

\newcommand{\jbe}{\mathtt{j}}
\newcommand{\rme}{\mathrm{e}}
\newcommand{\rmi}{\mathrm{i}}
\newcommand{\rmd}{\mathrm{d}}
\newcommand{\vecb}[1]{\boldsymbol{\mathrm{#1}}}
 
\newcommand{\eref}[1]{(\ref{#1})}

\newcommand{\Eref}[1]{Eq.~(\ref{#1})}
\newcommand{\Esref}[1]{Eqs.~(\ref{#1})}
\newcommand{\Sref}[1]{Sec.~\ref{#1}}
\newcommand{\Fref}[1]{Fig.~\ref{#1}}

\newcommand{\Hz}{\mathrm{Hz}}

\def\sig{{\tiny\textrm{sig}}}
\newcommand{\fkdot}[1]{f^{(#1)}}
\newcommand{\nukdot}[1]{\nu^{(#1)}}
\newcommand{\tnudot}[1]{\dot{\tilde{\nu}}}

\newcommand{\nud}{\dot{\nu}}
\newcommand{\nudd}{\ddot{\nu}}
\def\Doppler{\vecb{\mathrm{p}}}
\def\vDoppler{\Doppler}
\def\Xsym{\star}
\newcommand{\F}{\mathcal{F}}

\newcommand{\Xstat}{\F^\Xsym}
\newcommand{\XstatS}{\overline{\F}^\Xsym}

\newcommand{\Pspace}{\mathcal{P}}
\newcommand{\Psearch}{\mathcal{S}_\Pspace}
\newcommand{\xil}[1]{{\vec{\xi}}^{\,(#1)}}
\newcommand{\xic}{\vec{\xi}}
\newcommand{\ang}{\varphi}
\newcommand{\nx}{n_\mathrm{x}}
\newcommand{\ny}{n_\mathrm{y}}
\newcommand{\mm}{\mathcal{M}}
\newcommand{\mmS}{\overline{\mm}}
\def\av#1{{\langle #1 \rangle}}
\def\avj#1{{\langle #1 \rangle_{\tiny{[j]}} }}

\newcommand{\PhiErr}{\Delta\phi_{s}}
\newcommand{\Ntcoarse}{\mathcal{N}_t^{\tiny\textrm{coarse}}}
\newcommand{\Ntfine}{\mathcal{N}_t^{\tiny\textrm{fine}}}

\newtimeformat{dottime}{\twodigit{\THEHOUR}:\twodigit{\THEMINUTE}}
\settimeformat{dottime}

\DeclareGraphicsExtensions{.png}

 \def\dcc{LIGO-P1000006-v4}
 \def\aei{AEI-2010-036}

\begin{document}

\pagestyle{fancy}

\preprint{\dcc}
\preprint{\aei}

\rhead[]{}
\lhead[]{}

\title{Parameter-space metric of semicoherent searches for continuous gravitational waves}

\author{Holger J. Pletsch}
\email{Holger.Pletsch@aei.mpg.de} 
\affiliation{Max-Planck-Institut f\"ur Gravitationsphysik (Albert-Einstein-Institut)
    and Leibniz Universit\"at Hannover, Callinstra{\ss}e 38, D-30167 Hannover, Germany}

\date{\currenttime, \today}

\begin{abstract}
Continuous gravitational-wave (CW) signals such as emitted by spinning neutron stars are an important target class for current detectors. However, the enormous computational demand prohibits fully coherent broadband all-sky searches for prior unknown CW sources over wide ranges of parameter space and for yearlong observation times. More efficient hierarchical ``semicoherent" search strategies divide the data into segments much shorter than one year, which are analyzed coherently; then detection statistics from different segments are combined incoherently. To optimally perform the incoherent combination, understanding of the underlying parameter-space structure is requisite. This problem is addressed here by using new coordinates on the parameter space, which yield the first analytical parameter-space metric for the incoherent combination step. This semicoherent metric applies to broadband all-sky surveys (also embedding directed searches at fixed sky position) for isolated CW sources. Furthermore, the additional metric resolution attained through the combination of segments is studied. From the search parameters (sky position, frequency, and frequency derivatives), solely the metric resolution in the frequency derivatives is found to significantly increase with the number of segments.
\end{abstract}


\maketitle

\section{Introduction\label{sec:Introduction}}

Direct detection of gravitational waves would not only validate 
Einstein's theory of general relativity but also constitute an important new 
astronomical tool.
Continuous gravitational-wave (CW) signals are expected, for instance, from rapidly 
rotating neutron stars through various emission mechanisms 
\cite{owen-1998-58,ushomirsky:2000,cutler:2002-66,jones-2002-331,owen-2005-95}. 
Most such stars are anticipated to be electromagnetically 
invisible, but might be detected and studied via gravitational waves.
With current Earth-based detectors, such as LIGO~\cite{ligoS5},
numerous efforts are underway to search for CW sources 
\cite{pshS4:2008,S4EAH,S5R1EAH,eahurl}, and
observational upper limits have already allowed one to constrain 
the physics of neutron stars~\cite{crab-2008,powerflux:2009}. 

The expected CW signals are extremely weak. Hence
detection requires very sensitive data analysis techniques
to extract these signals from detector noise.  
In the work of~\cite{jks1} a powerful method has been derived
which is based on the principle of maximum likelihood detection,
leading to coherent matched filtering. The CW signals are
quasimonochromatic with a slowly changing intrinsic frequency.  
For a terrestrial detector, the Earth's motion relative to the solar
system barycenter (SSB) generates a Doppler modulation in amplitude and
phase of the waveform.  As shown in~\cite{jks1}, the parameters describing 
the signal's amplitude variation may be analytically eliminated by maximizing the
coherent matched-filtering statistic.  Thus, one only searches over
the remaining parameters describing the signal's phase:
the source's sky location, frequency, and frequency derivatives (``spindowns''). 
The so-obtained coherent detection statistic is called
the \mbox{$\F$-statistic}, which can also be generalized to employ
multiple detector data streams \cite{CutlerMultiIFO}.

Finite computational resources are what imposes severe limits on the sensitivity of
broadband all-sky searches for prior unknown CW sources~\cite{jks1,bccs1:1998}.
In the fully coherent \mbox{$\F$-statistic} approach, one must convolve the 
full data set with signal waveforms (templates) corresponding to all 
possible sources. But the number of templates required to cover
the search parameter space increases as a high power of the coherent integration time. 
For year-long data sets, searching a realistic range 
of parameter space would demand more computing power than
available on Earth~\cite{bccs1:1998,jks1}.  Therefore, a
fully coherent search is restricted to much shorter integration times.

A more efficient analysis of data sets which span yearlong time 
intervals is achieved via less expensive hierarchical semicoherent methods 
\cite{bc2:2000,cutler:2005,hough:2005,pletsch:GCT}.
In such a method, the data are divided into segments of duration~$T$, 
where~$T$ is much smaller than one year.  
This allows one to use a \emph{coarse} grid of templates,
on which the \mbox{$\F$-statistic} is calculated 
for each segment. Then the $\F$-statistics from all segments 
(or statistics derived from~$\F$) are incoherently combined using 
a common \emph{fine} grid of templates.
Since phase information is discarded between segments,
this latter step is called incoherent and thus the 
search methodology as a whole is referred to as \emph{semicoherent}.

An important long-standing problem in these semicoherent methods has been 
the design of, and link between, the coarse and fine grids. 
To address this problem it is essential to understand the underlying
parameter-space structure. In this context, the geometric approach
has been proven to be very useful,
introducing a \emph{metric} on parameter space, as  first done in
 \cite{Sathy1:1996,owen:1996me}.
The key quantity in this respect is called \emph{mismatch}~$\mm$,
which is the fractional loss in expected $\F$-statistic 
(or sum of expected \mbox{$\F$-statistics} in the incoherent step) 
for a given signal $\vDoppler_\sig$ at a nearby grid point~$\vDoppler$ in 
phase-parameter space~$\Pspace$.  Taylor-expanding $\mm$ (to quadratic order) 
at $\vDoppler_\sig$ in the differences of $\vDoppler_\sig$ and $\vDoppler$ defines the
parameter-space metric.

While the parameter-space metric for coherent searches has been
examined in detail \cite{bccs1:1998,ptolemetric,prix:2007mu},
the metric for semicoherent searches is comparably much less well studied.
However, the literature does exhibit several quests for a semicoherent metric
\cite{cutler:2005,hough:2005,Dhurandhar2008,Watts2008}.
About a decade ago, the first general discussion of the semicoherent 
metric for CW searches was given in~\cite{bc2:2000}, along with numerical
investigations. But only recently, the first fully analytical semicoherent
metric has been found, leading to a significantly improved CW search 
method~\cite{pletsch:GCT}. 
This recent progress is based on a better understanding of the global parameter-space 
correlations~\cite{pletsch:2008}, which were first examined in~\cite{prixitoh:2005}.
In turn, this insight provides new coordinates on parameter space, enabling the
analytical calculation and study of the semicoherent metric.

The present paper extends the recent work of~\cite{pletsch:GCT}
to greater generality and provides the essential technical basis for 
a parameter-space metric formalism for semicoherent CW searches. 
Additionally, complete and fully analytic semicoherent metric results 
are presented, which are ready to use for practical implementations
of semicoherent searches, serving the earlier-mentioned quests 
of previous literature.
The results apply to broadband all-sky surveys (embedding
directed searches with fixed sky position) for isolated CW sources.

Section~\ref{sec:MatchedFilter} briefly reviews matched filtering for CW signals 
and the parameter-space metric formalism in general.
Section~\ref{sec:SemiCohSearches} elucidates how the metric is obtained
for semicoherent searches. To evaluate and study the semicoherent metric,
new coordinates on parameter space are defined in \Sref{sec:NewCoords}.
In \Sref{sec:OneSpindown}, the semicoherent metric is derived and 
investigated for the case of CW signals whose intrinsic frequency
changes linearly with time (parametrized by one spindown parameter).
In particular, the refinement factor is introduced as the ratio of the number fine-grid 
and coarse-grid templates, quantifying
the additional parameter-space metric resolution due to
combination of many segments. 
Section~\ref{sec:TwoSpindowns} extends these results to 
the case of CW signals whose intrinsic frequency can
change quadratically with time (considering two spindown parameters).
Finally, a short conclusion follows in \Sref{sec:Conclusion}.

\section{Matched filtering for continuous gravitational-wave signals\label{sec:MatchedFilter}}

The  detector output data time series is denoted by $x(t)$ at detector time~$t$.
In the absence of any signal, the data contain only noise~$n(t)$, which is assumed to be a
zero-mean, stationary, and Gaussian random process. 
In case a signal~$h(t)$ is present, the noise is assumed to be additive, 
such that $x(t) = n(t) + h(t)$.  

The dimensionless signal response function~$h(t)$ of an 
interferometric detector to a weak plane gravitational wave in the 
long-wavelength approximation is
a linear combination of the form
\begin{equation}
  h(t) = F_+ (t) \, h_+(t) + F_\times (t) \, h_\times(t) ,
  \label{e:h_t}
\end{equation}
where $F_{+,\times}$ are called the beam-pattern functions,
resulting in the amplitude modulations from Earth's spinning motion.
They lie in the range $-1\le F_{+,\times}\le 1$,
and depend on the orientation of the detector and source, 
and on the polarization angle~$\psi$ of the waves. For explicit expressions
the reader is referred to~\cite{jks1}.
In the case of an isolated rapidly rotating neutron
star with nonaxisymmetric deformations and negligible proper motion (cf. \cite{jk2,jk3}), 
the waveforms corresponding to the plus~($+$) and cross~($\times$) 
polarizations are 
\begin{equation}
  h_+ (t) = A_+ \, \sin \Psi(t)\,, \qquad 
  h_\times (t) = A_\times \, \cos \Psi(t) \,,
\end{equation}
where $A_+$ and $A_\times$ are the plus and cross polarization
amplitude parameters, respectively, and $\Psi(t)$ is given by
\begin{align}
   \Psi(t) &= \Phi_0 + \Phi(t) \nonumber\\
   &= \Phi_0 + 2\pi  \sum_{k=0}^{s} \, \frac{\fkdot{k}(t_0)}{(k+1)!} 
  \,  \left[t-t_0 + \frac{\vec r (t) \cdot \vec n}{c}\right]^{k+1},
  \label{e:Phase1}
\end{align}
where $\Phi_0$ is the initial phase, $\fkdot{0} \equiv f$ denotes the 
frequency, and $\fkdot{k>0}$ is the $k$th frequency time derivative (also called ``spindown''), 
evaluated at the SSB at time $t_0$.  
The integer~$s>0$ denotes the number
of frequency time derivatives to be taken into account, 
therefore it holds $\fkdot{k>s}=0$. The vector $\vec r (t)$ connects from the SSB to the detector, 
 $c$ is the speed of light, and  $\vec n$ is a constant unit vector pointing
 from the SSB to the location of the CW source. 
Thus, a point in phase parameter space $\vDoppler \in \Pspace$ is denoted
by $\vDoppler=\left(\fkdot{k},\vec n\right)$ in respect of the reference time~$t_0$.

As first shown in~\cite{jks1}, the phase $\Phi(t)$ in \Eref{e:Phase1} can be 
approximated without significant loss in signal-to-noise ratio (SNR) to good
accuracy  by
\begin{eqnarray}
   \Phi(t) &\approx& 2\pi\,
   \sum_{k=0}^{s}\,\frac{\fkdot{k}(t_0)\;{(t-t_0)}^{k+1}}{(k+1)!}\nonumber\\
   & & + 2\pi\, \frac{\vec{r}(t)}{c}\cdot \vec{n}\; \sum_{k=0}^{s} \, \frac{\fkdot{k}(t_0)\;{(t-t_0)}^k}{k!}\,.
   \label{e:phase-model1}
\end{eqnarray}

Consider a data segment spanning the interval $[-T/2,T/2]$.
The $\F$-statistic is obtained~\cite{jks1,lrr-2005-3} from the likelihood 
ratio~$\Lambda$, which takes the form
\begin{equation}
  \ln \Lambda = (x|h) - \frac{1}{2} (h|h) \,,
  \label{e:loglikelihood1}
\end{equation}
where the scalar product has been defined as
\begin{equation}
  (x|y) \equiv 4\Re \int_{0}^{\infty}\, \frac{\tilde{x}(f)\,\tilde{y}^{\ast}(f)}{S_n(f)} \rmd f \,,
  \label{e:scalarpro1}
\end{equation}
with the Fourier transform indicated by $\tilde{}$, the complex
conjugation denoted by ${}^\ast$, and  $S_n$ defined as the one-sided noise
spectral density. One may assume $S_n$ 
to be constant over the narrow bandwidth of the signal considered in this
work. Then the scalar product of \Eref{e:scalarpro1} is approximately given by
\begin{equation}
  (x|y) \approx \frac{2}{S_n} \int_{-T/2}^{T/2}\, x(t)\,y(t) \rmd t \,.
  \label{e:scalarpro2}
\end{equation}
Thus, the time average is introduced by the following notation:
\begin{equation}
   \av{x} \equiv \frac{1}{T} \int_{-T/2}^{T/2}\, x(t)\, \rmd t \,.
\end{equation}
Using this notation, \Eref{e:loglikelihood1} is rewritten as
\begin{equation}
  \ln \Lambda = \frac{2 T}{S_n} \left[  \av{x\;h} - \frac{1}{2} \av{h^2}\right] \,.
  \label{e:loglikelihood2}
\end{equation}
Replacing the amplitude parameters $\{A_+,A_{\times},\psi,\Phi_0\}$ 
by their values which maximize $\ln \Lambda$, the so-called maximum 
likelihood (ML) estimators, defines the detection statistic~$\F$ as
\begin{equation}
  \F \equiv \ln \Lambda_{\rm ML}.
  \label{e:Fstatistic}
\end{equation}
Because the $\F$-statistic is maximized over the amplitude parameters, the 
remaining search space is just the phase-parameter space $\Pspace$.

\subsection{Coherent detection statistic for a simplified signal model\label{ssec:simplified}}

Since the primary goal of this work is in relation to template-grid construction,
a very useful approximation for this purpose is to replace
the beam-pattern functions $F_{+,\times}(t)$ by constant 
values \cite{jk2,jk4,pletsch:2008}.
The phase of the CW signal is expected to change very rapidly at the terrestrial detector site 
over a time scale of typically less than ten seconds, whereas the amplitude of 
the signal varies with a period of one sidereal day.  
Coherent observation times of practical interest are typically longer than
one day, so that replacing the beam-pattern functions $F_{+,\times}(t)$ with 
effective constant values is a good approximation.
In this case the signal model in \Eref{e:h_t} takes the form
\begin{equation}
     h(t) = A_1\, \sin \Phi(t) + A_2\, \cos \Phi(t) \,,
  \label{e:const-amplitudes}
\end{equation}
where $A_{1,2}$ are defined to be the constant amplitude parameters. 
In~\cite{jk2}, the validity of this approximation is also investigated using Monte Carlo 
simulations. It should be noted that the actual computation of the 
$\F$-statistic in a CW search will, of course, include the effects of amplitude 
modulation and involves precise calculation of the detector position with 
respect to the SSB using an accurate ephemeris model. This simplified
signal is used here to facilitate the template-grid construction.

The log likelihood of \Eref{e:loglikelihood2} 
for the simplified signal model \eref{e:const-amplitudes} is denoted by 
$\ln \Lambda^{\star}$ and takes the form
\begin{equation}
  \ln \Lambda^{\star} 
  = \frac{2 T}{S_n} \left[  A_1\,\av{x\;\sin\Phi} +   A_2\,\av{x\; \cos\Phi} - \frac{A_1^2+A_2^2}{4}\right] \,.
  \label{e:loglikelihood3}
\end{equation}
By substituting the ML estimators for $A_{1,2}$ 
in $\ln \Lambda^{\star}$ of \Eref{e:loglikelihood3}, it is straightforward
to show \cite{prixitoh:2005,pletsch:2008} that the simplified 
signal model \eref{e:const-amplitudes} leads to the following coherent detection 
statistic $\Xstat$ approximating $\F$ as
\begin{equation}
   \Xstat = \ln \Lambda^{\star}_{\rm ML} = \frac{2 T}{S_n}\,\left| \av{x\; \rme^{-i \Phi}} \right|^2 \,.
   \label{e:XstatDef}
\end{equation}

\subsection{Perfect match of signal and template phase parameters\label{ssec:known-signal}}

Consider a signal $h_\sig(t)$ following the model of \Eref{e:const-amplitudes} 
present in the data $x(t)$. Let the signal's phase evolution $\Phi_\sig(t)$ be described
by \emph{known} phase parameters denoted by the vector~$\vDoppler_\sig$ and 
defined at $t_0$,
\begin{equation}
  h_\sig(t) = A_{1,\sig}\, \sin \Phi_\sig(t) + A_{2,\sig}\, \cos \Phi_\sig(t) \,.
  \label{e:signalmodel1}
\end{equation}
Since the signal parameters
are known, one can construct a template which perfectly matches the signal.
Assuming the noise $n(t)$ to be stationary, Gaussian, zero-mean, and additive,
one can show~\cite{jk5,jkbook} that for a known signal (zero parameter offsets) 
the expectation value and variance of $\Xstat$, respectively, are
\begin{equation}
  E\left[ \Xstat \right]  = 1 + \frac{1}{2}\, \rho^2(0) \,,\quad \sigma^2_{\tiny{\Xstat}} = 1+ \rho^2(0)\,,
\end{equation}
where $\rho(0)$ defines the optimal SNR, obtained as
\begin{equation}
  \rho^2(0) =  \frac{4 T}{S_n} \,\left| \av{h_\sig \; \rme^{-i \Phi_\sig}} \right|^2 \,.
  \label{e:optSNR}
\end{equation}
The expression of \Eref{e:optSNR} may be further simplified using \Eref{e:signalmodel1}
to yield
\begin{equation}
  \rho^2(0) =  \frac{T}{S_n} \,\left( A_{1,\sig}^2 + A_{2,\sig}^2 \right) \,.
  \label{e:optSNR2}
\end{equation}

\subsection{Mismatch of the signal and template phase parameters \label{ssec:unknown-signal}}

If the signal parameters are \emph{unknown} in advance, one 
has to evaluate the detection statistic for a bank of templates.
Let the template phase-parameter vector be~$\vDoppler$
and the corresponding phase be $\Phi(t)$.
In general, the template phase parameters will not exactly match
the signal parameters. Thus, the parameter offsets are labeled by
\begin{equation}
  \Delta\vDoppler \equiv \vDoppler_\sig - \vDoppler \,.
\end{equation}
The resulting difference in phase $\Delta\Phi(t)$ between 
the phase $\Phi_\sig(t)$ of the signal and the phase $\Phi(t)$ of a 
template is defined as
\begin{equation}
  \Delta\Phi(t) \equiv \Phi_\sig(t)- \Phi(t) \,.
\end{equation}
In this case the expectation value and variance of $\Xstat$, respectively, are given by
\begin{equation}
  E\left[ \Xstat \right] = 1 + \frac{1}{2}\,\rho^2(\Delta\vDoppler)  \,,
  \quad \sigma^2_{\tiny{\Xstat}} = 1 + \rho^2(\Delta\vDoppler)\,
\end{equation}
where the SNR $\rho(\Delta\vDoppler)$ here depends on 
the parameter offsets~$\Delta\vDoppler$, such that
\begin{equation}
  \rho^2(\Delta\vDoppler) =  \frac{4 T}{S_n} \,\left| \av{h_\sig \; \rme^{-i \Phi} } \right|^2 \,.
  \label{e:mismatchedSNR}
\end{equation}
Further simplification of \Eref{e:mismatchedSNR} leads to
\begin{eqnarray}
  \rho^2(\Delta\vDoppler) &=&  \frac{T}{S_n} \,\left( A_{1,\sig}^2 + A_{2,\sig}^2 \right) \,
   \left| \av{\rme^{\rmi \Delta\Phi}} \right|^2 \nonumber\\
  &=& \rho^2(0) \,   \left| \av{\rme^{\rmi \Delta\Phi}} \right|^2 \,.
  \label{e:mismatchedSNR2}
\end{eqnarray}
The above relation shows that given an offset $\Delta\vDoppler$ between
the signal and template parameters the squared SNR is reduced by
$\left| \av{\rme^{\rmi \Delta\Phi}} \right|^2$. 
This gives rise to define a dimensionless  ``mismatch'' $\mm$ as
\begin{subequations}
\label{e:mm1}
\begin{align}
  \mm &= \frac{\rho^2(0) -  \rho^2(\Delta\vDoppler)}{\rho^2(0)} \\
      &= 1 - \left| \av{\rme^{\rmi \Delta\Phi} } \right|^2 \,.
\end{align}
\end{subequations}
The mismatch represents the fractional loss in the expected 
detection statistic due to the parameter~$\Delta\vDoppler$ and thus
provides a distance measure in the template parameter space.

\subsection{Metric on parameter space \label{ssec:metric1}}

Taylor-expanding the mismatch $\mm$ up to quadratic order 
in terms of the template parameter-space location offsets $\vDoppler$ 
at the signal location $\vDoppler_\sig$ yields
\begin{equation}
  \mm \approx \sum_{a,b} g_{ab}\,\Delta\vDoppler^a \,\Delta\vDoppler^b  \,,
  \label{e:mm-metric}
\end{equation}
defining a positive definite metric tensor $g$ as
\begin{equation}
   g_{ab} = \av{\partial_a \Phi\; \partial_b \Phi} 
   - \av{\partial_a \Phi}\av{\partial_b \Phi} \,,
 \label{e:metric-tensor1}
\end{equation}
where $a$ and $b$ label the tensor indices,
 and the following notation has been employed:
\begin{equation}
   \partial_a \Phi \equiv \left. \frac{\partial\Phi}{\partial\vDoppler^a}  \right|_{
  \vDoppler=\vDoppler_\sig} \,.
\end{equation}
The expression of \Eref{e:metric-tensor1} 
is often called the ``phase metric" \cite{bccs1:1998,jk3,prix:2007mu}, 
since it describes a distance measure 
on the phase parameter space~$\Pspace$.

\section{Parameter-space metric for semicoherent searches\label{sec:SemiCohSearches}}

In semicoherent CW searches the data are divided into $N$ segments
of duration $T$, where~$T$ is much smaller than one year.  
This allows to analyze each segment coherently,
using a \emph{coarse} grid of templates. 
Then the coherent detection statistic results from all segments are
incoherently combined using a common \emph{fine} grid of templates.
This scheme is often called  a semicoherent search strategy,
offering the best overall sensitivity at limited computational
resources \cite{bc2:2000,cutler:2005} when the fully coherent 
approach is infeasible.
In preparation of calculating the semicoherent metric, this section
introduces some general notation.

\subsection{The coherent metric for a given segment \label{ssec:CohMetric}}

Let the integer $j=1,...,N$ label the $j$th segment, and let~$t_j$ denote the 
time midpoint of segment~$j$.
The time average over the $j$th segment is defined by
\begin{equation}
  \avj{x} \equiv \frac{1}{T} \int_{t_j-T/2}^{t_j+T/2} x(t)\, \rmd t \,.
\end{equation}
The mismatch $\mm_j$ in the $j$th segment is given by
\begin{equation}
  \mm_j = \frac{\rho^2(0) -  \rho^2_j(\Delta\vDoppler)}{\rho^2(0)} \,,
  \label{e:mm1}
\end{equation}
where
\begin{equation}
   \rho^2_j(\Delta\vDoppler) = \rho^2(0) \, \left| \avj{\rme^{\rmi \Delta\Phi}} \right|^2 \,.
\end{equation}
Thus $\mm_j$ can be approximated by
\begin{equation}
  \mm_j \approx \sum_{a,b} g^{[j]}_{ab}\,\Delta\vDoppler^a \,\Delta\vDoppler^b  \,,
  \label{e:mm-metric-j}
\end{equation}
where the components of coherent metric tensor for the $j$th segment~$g^{[j]}$
are obtained in analogy to \Eref{e:metric-tensor1} as
\begin{equation}
  g^{[j]}_{ab} = \avj{\partial_a\Phi\; \partial_b\Phi} 
 - \avj{\partial_a\Phi}\avj{\partial_b\Phi} \,.
 \label{e:metric-tensor2-j}
\end{equation}

When searching a subspace~$\Psearch$ of the phase parameter space~$\Pspace$,
the corresponding proper volume $V$  is given by
\begin{equation}
  V = \int_{\Psearch} \rmd V = \int_{\Psearch} \sqrt{\det g^{[j]}}\, \rmd \vDoppler \,.
\end{equation}
The placement of signal templates to cover the search parameter space 
is an instance of the sphere covering problem~\cite{prix:2007ks}.
Using a lattice of templates, the number of coarse-grid templates 
$\Ntcoarse$ is obtained from the coherent metric tensor $g^{[j]}$ as
\begin{equation}
  \Ntcoarse =  \rho_0\,V =   \rho_0\,\int_{\Psearch} \sqrt{\det g^{[j]}}\, \rmd \vDoppler \,,
  \label{e:Ntcoarse}
\end{equation}
where the constant $\rho_0$ describes the density of templates. 
The specific value of  $\rho_0$ depends on the
desired maximum mismatch and the chosen type of lattice~\cite{bccs1:1998,prix:2007ks,jk5}.
When using a random template bank instead of a lattice,
then~$\rho_0$ can also depend on the desired 
coverage fraction~\cite{MessengerPrixPapa2009,MancaVallisneri2010}.

\subsection{The semicoherent metric for combining segments \label{ssec:SemiCohMetric}}

In the semicoherent search approach, coherent detection statistic results 
from the different segments are incoherently combined.
To evaluate the metric for this case,
the simplified coherent detection statistic $\Xstat$ (approximating $\F$)
of \Eref{e:XstatDef} is again used here. Thus, $\Xstat_j$ means 
the $\Xstat$-statistic value 
obtained in the $j$th segment. Recall that $\Xstat_j$ is the log likelihood function
(maximized over the amplitude parameters). As the joint likelihood is
the product, the joint log likelihood of all segments is the sum over $j$.
Therefore, we define the semicoherent detection statistic $\XstatS$ by
\begin{equation}
  \XstatS  = \frac{1}{N}\,\sum_{j=1}^N  \, \Xstat_j \,.
\end{equation}

For the case of a known signal (zero parameter offsets) 
the expectation value and variance of~$\XstatS$, respectively, are
\begin{equation}
  E\left[ \XstatS \right] = 1 + \frac{1}{2}\,\rho^2(0) \,,\quad \sigma^2_{\tiny{\XstatS}} = \frac{1 + \rho^2(0)}{N}\,,
\end{equation}
assuming identical noise spectral densities $S_n$ in every segment.
Hence, combining detection statistics from the $N$ segments
reduces the variance by~$N$.

For nonzero offsets~$\Delta\vDoppler$ between the template
and signal parameters the resulting expectation value and 
variance of~$\XstatS$, respectively, are obtained as
\begin{align}
  E \left[ \XstatS \right] &= 1 + \frac{1}{2N}\,\sum_{j=1}^N \rho^2_j(\Delta\vDoppler) \,,\\
  \quad \sigma^2_{\tiny{\XstatS}} 
  &= \frac{1 + \frac{1}{N}\,\sum_{j=1}^N \rho^2_j(\Delta\vDoppler)}{N}\,.
\end{align}

Thus, the mismatch~$\mmS$, which measures
the fractional loss in the expected semicoherent 
detection statistic $\XstatS$ due to phase-parameter offsets $\Delta\vDoppler$
is obtained as
\begin{align}
   \mmS &= \frac{1}{N}\,\sum_{j=1}^N \frac{\rho^2(0) -  \rho^2_j(\Delta\vDoppler)}{\rho^2(0)} \nonumber\\
   &=\frac{1}{N} \sum_{j=1}^{N}\, \mm_j\,.
\end{align}
Since $\mm_j$ is the mismatch in segment $j$ according to \Eref{e:mm-metric-j},
$\mmS$ represents the average mismatch across the segments~(cf. \cite{bc2:2000}).
Consequently, one may write
\begin{equation}
   \mmS \approx \sum_{a,b} \bar{g}_{ab}\,\Delta\vDoppler^a \,\Delta\vDoppler^b \,,
\end{equation}
where the components of the semicoherent metric tensor~$\bar{g}$ are obtained as
the average of the individual-segment coherent metric components 
$g^{[j]}_{ab}$ from \Eref{e:metric-tensor2-j}, 
\begin{equation}
  \bar{g}_{ab} = \frac{1}{N} \sum_{j=1}^N\, g^{[j]}_{ab} \,.
  \label{e:avegab}
\end{equation}

Thus, in analogy to \Eref{e:Ntcoarse} the number of fine-grid templates~$\Ntfine$
is given by
\begin{equation}
  \Ntfine =  \rho_0\, \int_{\Psearch} \sqrt{\det \bar{g}}\, \rmd \vDoppler \,.
  \label{e:Ntfine}
\end{equation}

\section{New coordinates on parameter space\label{sec:NewCoords}}

The standard ``physical'' coordinates on $\Pspace$ are the frequency
and frequency derivatives~$\fkdot{k}(t_0)$ at reference time $t_0$, and the unit
vector \mbox{$\vec n = (\cos \delta \, \cos \alpha, \cos \delta \, \sin
\alpha, \sin \delta)$} on the two-sphere~$S^2$, pointing from the SSB
to the source.  The angles $\alpha$ and $\delta$ are right ascension and
declination.    

The analytic evaluation of the semicoherent metric components 
$\bar{g}_{ab}$ from \Eref{e:avegab} 
is one of the central aspects of this work. This problem is approached 
by introducing new coordinates on the phase parameter space~$\Pspace$,
leading to a phase model which depends linearly on the coordinates.

\subsection{Linear phase model\label{ssec:LinPhaseModel}}

For coherent segment lengths~$T$ much smaller than one year,
the orbital component $\vec r_{\textrm{orb}}$ of the Earth's motion,
\mbox{$\vec r = \vec r_{\textrm{orb}} + \vec r_{\textrm{spin}}$}, 
varies slowly during~$T$ and thus can be Taylor expanded
around the segment's midpoint. Hence, by separating the orbital
and spinning components of the Earth's motion in the phase 
model $\Phi(t)$ a convenient reparameterization is obtained in which
$\Phi(t)$ depends linearly on the new coordinates. For further details
the reader is referred to Ref.~\cite{pletsch:2008}.
Thus the resulting phase model $\Phi(t)$ is obtained as
\begin{align}
   \Phi(t) =\, &\phi_0(t_0) 
   + \sum_{k=0}^s  \nukdot{k}(t_0) \,\left(\frac{t-t_0}{T}\right)^{k+1} 2^{k+1}\nonumber\\
   &+ \nx(t_0)\,\cos \Omega\,t + \ny(t_0)\,\sin \Omega\,t \,,
   \label{e:phase-new-coords}
\end{align}
where $\phi_0(t_0)$ is a constant independent of $t$, and
the new frequency and frequency-derivative coordinates $\nukdot{k}(t_0)$
as first derived in~\cite{pletsch:2008} are
\begin{align}
  {\nukdot{k}(t_0)} &\equiv  2\pi \left(\frac{T}{2}\right)^{k+1} \biggl[ \frac{\fkdot{k}(t_0)}{(k+1)!} \nonumber\\
  &\;\;\;\; +  \sum_{\ell=0}^{k+1}  
  \frac{\fkdot{\ell}(t_0)}{\ell! (k-\ell+1)!} \, \xil{k-\ell+1}(t_0) \cdot \vec{n}  \biggr]   \,,
  \label{e:C1}
\end{align}
and the new sky coordinates (as in \cite{jk4,pletsch:GCT}) are given by
\begin{subequations}
\label{e:skycoords}
\begin{align}
  \nx(t_0) &\equiv 2\pi  f(t_0) \, \tau_E \, \cos \delta_D\, \cos \delta \, \cos[\alpha - \alpha_D(t_0)] \,, \\
  \ny(t_0) &\equiv 2\pi  f(t_0)\, \tau_E \, \cos \delta_D\, \cos \delta \, \sin [\alpha - \alpha_D(t_0)] \,.
\end{align}
\end{subequations}
Thereby, $\xic(t)\equiv\vec{r}_{\rm orb}(t)/c$, with 
$\vec{r}_{\rm orb}(t)$ denoting the vector from the Earth's barycenter to the SSB,
and, \mbox{$\tau_E = R_E/c \approx 21\,\textrm{ms}$} is the light travel time
from the Earth's center to the detector, $\alpha_D(t_0)$, $\delta_D$
are the detector position at time $t_0$, and $\Omega=2\pi/(1\,\textrm{sd})$ is the 
angular velocity of the Earth's spinning motion, which has a period of one
sidereal day.

Apart from an overall factor, the quantities $\nukdot{k}$ have been referred to as
the ``global-correlation parameters''~\cite{pletsch:2008,pletsch:GCT}.
As similarly done in earlier work \cite{jk4,prix:2007mu}, 
note that  the parameters $\nukdot{k}$ include a rescaling factor 
of $(T/2)^{k+1}$, such that they become dimensionless for convenience.

\subsection{Validity estimation\label{ssec:ValEst}}

\begin{figure}[b]
  \includegraphics[width=0.9\columnwidth]{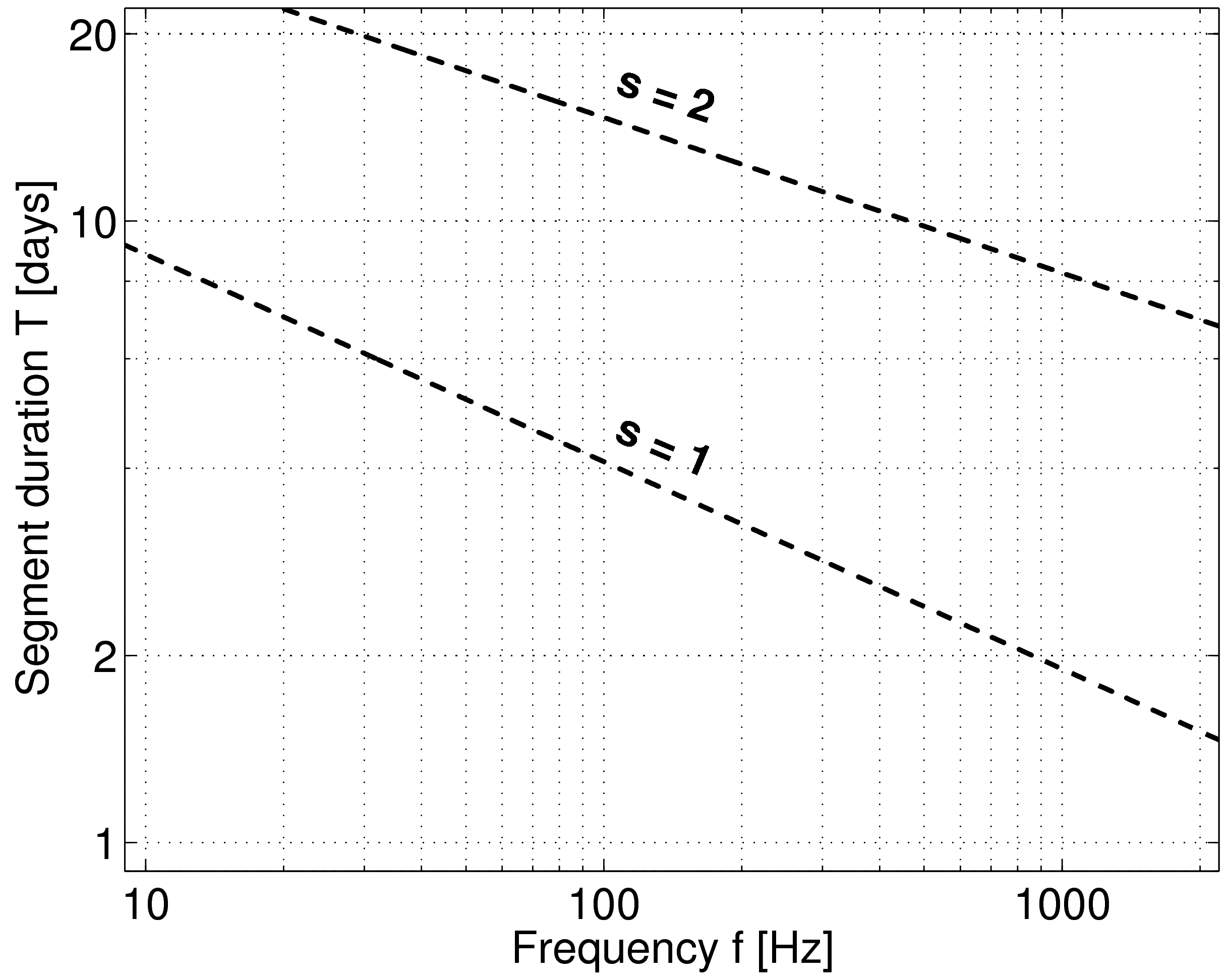}
  \caption{Validity estimation of the approximate phase model \Eref{e:phase-new-coords}
  in terms of the number of spindown parameters $s$ considered, for given
  values of coherent integration time $T$ and 
  the source frequencies searched. The dashed lines correspond to $30\%$
  mismatch for the one and two spindown case, respectively.
  For example, when including one spindown 
  parameter ($s=1$), the mismatch due to the approximate phase model 
  should be less than $30\%$ for coherent integrations~$T$
  up to about $2$~days while searching frequencies up to about~$1\textrm{k}\Hz$.
  \label{f:s-params-val} }
\end{figure}

The phase model of \Eref{e:phase-new-coords} is an approximation to
the phase evolution described by \Eref{e:phase-model1}.
The validity of this approximation depends on the coherent 
integration time $T$ (duration of a given segment) and on the source frequencies searched,
as was previously investigated in \cite{jk4,pletsch:2008}.
The maximum value of $T$ as a function of the highest search frequency $f$ 
may be estimated by considering the first neglected term in the phase model
becomes large enough to eventually lead to a significant mismatch.
Given the phase model of \Eref{e:phase-new-coords} and searching
$s$ spindown parameters, the first neglected term $\PhiErr $ is given by
\begin{align}
     & \PhiErr(t) \equiv 2\pi \frac{(t-t_0)^{s+2}}{(s+2)!}\,f(t_0) \, \xil{s+2}(t_0) \cdot \vec{n} \\
     & \hspace{1cm} \leq  2\pi \frac{(t-t_0)^{s+2}}{(s+2)!} \, f(t_0) \, \left|\xil{s+2}(t_0)\right| \,.
\end{align}
The mismatch produced by the phase offset $\PhiErr$ follows from 
\Eref{e:mm1} as $1 - \left| \av{\rme^{\rmi \PhiErr} } \right|^2$.
Figure~\ref{f:s-params-val} shows the values of~$T$ as a function of
frequency~$f$ for which $\PhiErr$ yields a mismatch of~$30\%$.
For instance, with $s=1$, one should be able to use coherent segment durations~$T$
up to about $2$~days for search frequencies~$f$ up to about $1\textrm{k}\Hz$,
such that the mismatch due to the approximate phase model of \Eref{e:phase-new-coords} 
is still less than about $30\%$.
These results also qualitatively agree with the earlier investigations reported 
in~\cite{jk4}. 

However, most importantly, the new coordinates $\{\nukdot{k},\nx,\ny\}$ 
have significant advantages over the original coordinates $\{\fkdot{k}, {\vec n}\}$.
These new coordinates permit the first analytical solution 
for the metric of the incoherent combination step in hierarchical searches, 
yielding the ``semicoherent metric'' to be presented in the following.
In addition, in these new coordinates, the obtained metric 
is explicitly coordinate independent, a very convenient
feature when it comes to practical aspects of conducting CW searches.

\section{Metric evaluation for one spindown parameter \label{sec:OneSpindown}}

For all-sky surveys of prior unknown CW sources the search
parameter space $\Psearch$ is typically a four-dimensional 
subspace of $\Pspace$ \cite{pshS4:2008,S4EAH,powerflux:2009,S5R1EAH},
restricting to linear changes in frequency (one spindown parameter). 
In this case, using the new coordinates a point in $\Pspace$
is labeled by the vector \mbox{$\vDoppler=\left(\nu,\nud,\nx,\ny \right)$}
at a given reference time $t_0$.
The phase model of \Eref{e:phase-new-coords}  with $s=1$ takes the form
\begin{align}
   \Phi(t) &= \phi_0(t_0) 
   +  \nu(t_0) \,\frac{2\,(t-t_0)}{T}
   + \nud(t_0) \,\frac{4\,(t-t_0)^2}{T^2} \nonumber\\
   &+ \nx(t_0)\,\cos \Omega\,t + \ny(t_0)\,\sin \Omega\,t \,,
   \label{e:phase-new-coords-1sd}
\end{align}
where  the coordinates $\nu$ and $\nud$ from \Eref{e:C1} are explicitly written as
\begin{subequations}
\label{e:OneSdNus}
\begin{equation}
\nu(t_0) = 2\pi \, \frac{T}{2} \biggl[ f(t_0) + f(t_0) \, \dot{\vec{\xi}}(t_0) \cdot \vec n
  + \dot{f}  \,\vec{\xi}(t_0) \cdot \vec n \biggr] \,, 
\end{equation}
\begin{equation}
\nud(t_0) = 2\pi \left(\frac{T}{2}\right)^2\biggl[ \frac{\dot{f}}{2} + \frac{f(t_0)}{2} \, \ddot{\vec{\xi}}(t_0) \cdot \vec n 
  + \dot{f} \,  \dot{\vec{\xi}}(t_0) \cdot \vec n \biggr]  \,.
\end{equation}
\end{subequations}
The coordinates $\nx$ and $\ny$ are as given by \Esref{e:skycoords}.

\subsection{Coherent metric\label{ssec:OneSpindownCohMet}}

Using the coordinates $\{\nu,\nud,\nx,\ny \}$, the
components $g^{[j]}_{ab}$ of the symmetric coherent metric tensor  can be
computed analytically from \Eref{e:metric-tensor2-j}. 
Thereby it is useful to define the dimensionless quantity $\ang$ as
\begin{equation}
  \ang\equiv \Omega\,\frac{T}{2} \,.
\end{equation}
Thus from \Eref{e:metric-tensor2-j} the components  $g^{[j]}_{ab}$  
are obtained as
\begin{subequations}\label{e:OneSdCohMetricComps1}
\begin{align}
g^{[j]}_{\nu\nu} =&\, \frac{1}{3} \,, \\
g^{[j]}_{\nu\nud} =&\, \frac{4}{3} \; \left(\frac{t_j -t_0}{T} \right)\,,\\
g^{[j]}_{\nud\nud} =&\, \frac{4}{45} + \frac{16}{3}\; \left(\frac{t_j -t_0}{T}\right)^2  \,,\\
g^{[j]}_{\nu\nx} =&\, -\jbe_1(\ang) \; \sin \left(\Omega\,t_j\right) \,,\\
g^{[j]}_{\nu\ny} =&\,  \jbe_1(\ang) \; \cos \left(\Omega\,t_j\right)\,,\\
g^{[j]}_{\nud\nx} =&\, - \frac{2}{3}\;\jbe_2(\ang)\; \cos \left(\Omega\,t_j \right) \nonumber\\
&\;- 4\,\jbe_1(\ang) \; \left(\frac{t_j-t_0}{T}\right)  \sin \left(\Omega\,t_j \right) \,,\\
g^{[j]}_{\nud\ny} =& - \frac{2}{3}\;\jbe_2(\ang)\; \sin \left(\Omega\,t_j \right)\nonumber\\
&\; +4\, \jbe_1(\ang)\;\left(\frac{t_j-t_0}{T}\right) \cos \left(\Omega\,t_j \right) \,,
\end{align}
\begin{align}
g^{[j]}_{\nx\nx} =&\, \frac{1}{2} - \frac{1}{2} \jbe_0(\ang)\, \cos \left(\ang\right)
- \jbe_1(\ang)\, \sin \left(\ang\right) \, \cos^2 \left(\Omega\,t_j \right) \,,  \\
g^{[j]}_{\nx\ny} =&\,  -\jbe_1(\ang)\, \sin \left(\ang\right) \,
\sin \left(\Omega\,t_j\right)\,  \cos \left(\Omega\,t_j\right) \,, \\
g^{[j]}_{\ny\ny} =&\, \frac{1}{2} - \frac{1}{2} \jbe_0(\ang)\, \cos \left(\ang\right)
- \jbe_1(\ang)\, \sin \left(\ang\right) \, \sin^2 \left(\Omega\,t_j \right) \,,
\end{align}
\end{subequations}
where the spherical Bessel functions $\jbe_n(x)$  \cite{stegun} are  defined by
\begin{equation}
   \jbe_n(x) \equiv (-x)^n\,\left(\frac{1}{x}\,\frac{\rmd}{\rmd x}\right)^n \, \frac{\sin(x)}{x}\,.
   \label{e:Besselj}
\end{equation}
The first few spherical Bessel functions are given by
\begin{subequations}
\begin{align}
  \jbe_0(x) &= \frac{\sin(x)}{x} \,,\\
  \jbe_1(x) &= \frac{\sin(x)}{x^2} - \frac{\cos(x)}{x} \,,\\
  \jbe_2(x) &= \left[\frac{3}{x^2} - 1 \right]\,\frac{\sin(x)}{x} - \frac{3\,\cos(x)}{x^2}\,,\\
  \jbe_3(x) &= \left[\frac{15}{x^3} - \frac{6}{x} \right]\,\frac{\sin(x)}{x} 
     -  \left[\frac{15}{x^2} - 1 \right]\,\frac{\cos(x)}{x}\,.
\end{align}
\end{subequations}
Note that the components $g^{[j]}_{ab}$ of the coherent metric tensor  are
\emph{explicitly independent} of the coordinates. Therefore,
the number of coarse-grid templates $\Ntcoarse$ as described
by \Eref{e:Ntcoarse} can be rewritten as
\begin{equation}
  \Ntcoarse =  \rho_0\, \sqrt{\det g^{[j]}} \int_{\Psearch}  \rmd \vDoppler \,,
  \label{e:Ntcoarse2}
\end{equation}
where $\sqrt{\det g^{[j]}}$ has been taken outside the integration
over the searched region of parameter space, since it is independent of the coordinates. 
Thus, $\sqrt{\det g^{[j]}}$ directly scales the number of templates~$\Ntcoarse$. 
The actual value of $\Ntcoarse$ depends
on the parameter-space region $\Psearch$ searched over. 
To analytically obtain realistic estimates for $\Ntcoarse$, one may assume the ranges
\begin{align}
   \pi T f_{\tiny \textrm{min}} \lesssim\; &\nu \lesssim \pi T f_{\tiny \textrm{max}} \,,\label{e:range-nu}\\
   -\pi\frac{T^2\,f}{4\,\tau_{\tiny \textrm{min}}}  \lesssim\; &\nud \lesssim \pi\frac{T^2\,f}{4\,\tau_{\tiny \textrm{min}}} \,,    \label{e:range-nud}
\end{align}
where $\tau_{\tiny \textrm{min}}=f/\dot f$ represents the ``minimum spindown age''~\cite{jks1}
to search for.
At fixed frequency~$f$ the parameter ranges of $\nx$ and $\ny$ 
determine a two-dimensional disk $\mathcal{D}_f$ with radius of about $2\pi f \tau_E$.
Thus, \Eref{e:Ntcoarse2} yields
\begin{align}
 \Ntcoarse &=  \rho_0 \sqrt{\det g^{[j]}} \nonumber\\
  &\hspace{0.5cm}\times\int_{\pi T f_{\tiny \textrm{min}}}^{\pi T f_{\tiny \textrm{max}}}  \rmd \nu 
  \int_{\mathcal{D}_f} \rmd\nx\, \rmd\ny 
  \int_{-\frac{\pi T^2 f}{4\,\tau_{\tiny \textrm{min}}}}^{\frac{\pi T^2 f}{4\,\tau_{\tiny \textrm{min}}}} 
   \rmd \nud\nonumber\\
 &\approx  \rho_0 \sqrt{\det g^{[j]}}\,\frac{\pi^5 \tau_E^2}{2\,\tau_{\tiny \textrm{min}}} \,
  T^3 \left( f^4_{\tiny \textrm{max}} -f^4_{\tiny \textrm{min}}\right)\,.
  \label{e:Ntcoarse3}
\end{align}

The determinant of the coherent metric tensor $\det g^{[j]}$ is 
obtained as
\begin{align}
 \det g^{[j]} &= \frac{1}{135} \; 
 \bigl[1 -6\,{\jbe_1}^2(\ang) - {\jbe_0}(\ang)\,\cos\left(\ang\right) \bigr] \nonumber\\
 &\hspace{0.5cm}\times \bigl[ 1 - 10\,{\jbe_2}^2(\ang) - {\jbe_1}(\ang)\,\sin\left(\ang\right) \nonumber\\
 &\hspace{1cm}- {\jbe_0}(\ang)\,\cos\left(\ang\right) \bigr] \,.
   \label{e:detgabj0}
\end{align}
Note that $\det g^{[j]}$ is also explicitly independent of the
reference time $t_0$ (as well as of $t_j$), 
and solely depends upon $\ang$. This can be understood from
the following reasoning. For the parameters $\fkdot{k}$ a change of
reference time corresponds to a linear transformation~\cite{jk2,jk3},
whose determinant is~$1$. Similarly, for the parameters
$\nx$ and $\ny$ a change of reference time can be
described by a rotation, whose determinant is also~$1$.
Therefore this explains why $\det g^{[j]}$ is independent
of the reference time.

\begin{figure}[t]
  \includegraphics[width=0.9\columnwidth]{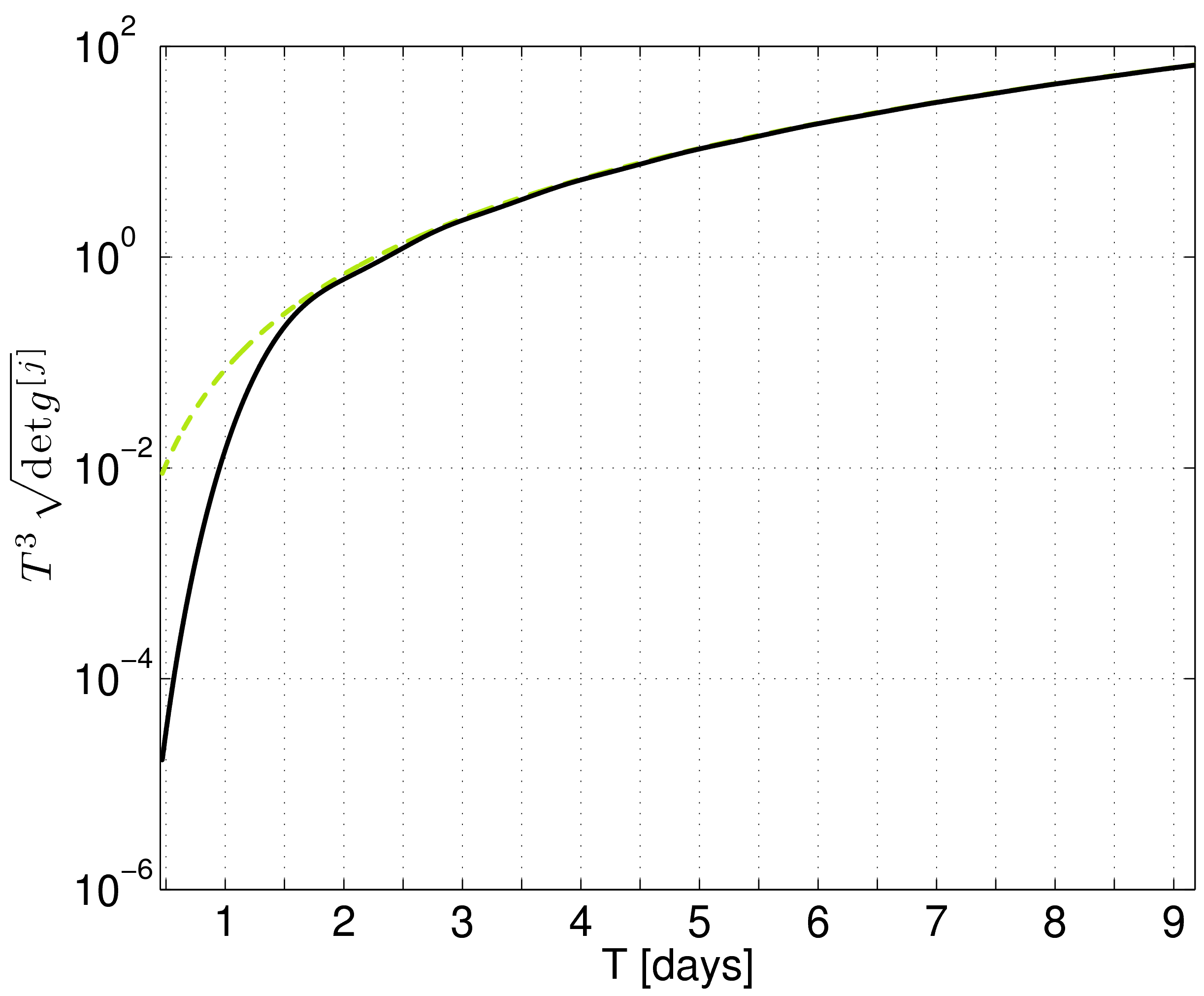}
  \caption{
  Dependency of the number of coarse-grid templates~$\Ntcoarse$
  on the coherent integration time $T$ (segment length) for the
  one-spindown case.
  The solid curve shows $T^3\,\sqrt{\det g^{[j]}}$, since
  \mbox{$\Ntcoarse \propto T^3\,\sqrt{\det g^{[j]}}$}.
  The dashed curve represents $T^3/\sqrt{135}$.
  \label{f:detgabj0-s1-SD1} }
\end{figure}

To examine the scaling of $\Ntcoarse$ with the coherent integration time~$T$,
\Fref{f:detgabj0-s1-SD1} shows $T^3\,\sqrt{\det g^{[j]}}$,
which is a measure for~$\Ntcoarse$, as  a function of~$T$. 
For increasing values of $T$, $\det g^{[j]}$ converges to $1/135$, 
since the metric tensor components related to the Earth's spinning
motion, $\nx$ and $\ny$, become approximately constant.
This behavior can be understood from the Rayleigh criterion.
An estimate of the diffraction-limited resolution is described 
by the ratio of the wavelength and the 
effective ``baseline''~\cite{prix:2007mu}. The maximum baseline 
in terms of the Earth's spinning motion is the Earth's diameter, 
which is first reached already after half a day of integration. 
Therefore, very little metric resolution is gained after 
integration times $T$ of about a day,  
as long as the Earth's orbital motion can still 
be well modeled by a Taylor expansion.

Finally, it should be mentioned that a convenient choice of~$t_0$
in favor of a compact notation is \mbox{$t_0=t_j=0$}.
To indicate this choice has been made, the resulting coherent metric 
tensor is denoted by $g^{[j=0]}$. 
The components $g^{[j=0]}_{ab}$ are obtained from 
\Esref{e:OneSdCohMetricComps1} as follows:
\begin{subequations}\label{e:OneSdCohMetricComps2}
\begin{align}
g^{[j=0]}_{\nu\nu} &= \frac{1}{3} \,, \\
g^{[j=0]}_{\nu\nud} &= 0 \,,\\
g^{[j=0]}_{\nud\nud} &= \frac{4}{45} \,, \\
g^{[j=0]}_{\nu\nx} &=  0\,,\\
g^{[j=0]}_{\nu\ny} &=  \jbe_1(\ang) \,,\\
g^{[j=0]}_{\nud\nx} &=  - \frac{2}{3}\;\jbe_2(\ang) \,, \\
g^{[j=0]}_{\nud\ny} &= 0 \,, \\
g^{[j=0]}_{\nx\nx} &=\frac{1}{2} - \frac{1}{2} \jbe_0(\ang)\, \cos \left(\ang\right)
- \jbe_1(\ang)\, \sin \left(\ang\right) \,, \\
g^{[j=0]}_{\nx\ny} &= 0 \,, \\
g^{[j=0]}_{\ny\ny} &=  \frac{1}{2} - \frac{1}{2} \jbe_0(\ang)\, \cos \left(\ang\right) \,.
\end{align}
\end{subequations}

\subsection{Semicoherent metric\label{sssec:OneSpindownSemiCohMet}}

Given the coherent metric tensor 
components~$g^{[j]}_{ab}$ in \Esref{e:OneSdCohMetricComps1}, 
explicit expressions for the components $\bar{g}_{ab}$ of the semicoherent 
metric tensor~$\bar{g}$ are obtained via \Eref{e:avegab} as
\begin{subequations}\label{e:OneSdSemiCohMetricComps1}
\begin{align}
\bar{g}_{\nu\nu} =&\, \frac{1}{3} \,, \\
\bar{g}_{\nu\nud} =&\, \frac{4}{3}\,\mu_1  \,, \\
\bar{g}_{\nud\nud} =&\, \frac{4}{45} + \frac{16}{3}\,\mu_2  \,, \\
\bar{g}_{\nu\nx} =&\,- \jbe_1(\ang)\,  \mu^{\textrm{\tiny SIN}}_0 \,, \\
\bar{g}_{\nu\ny} =&\, \jbe_1(\ang)\,  \mu^{\textrm{\tiny COS}}_0 \,, \\
\bar{g}_{\nud\nx} =&\, - \frac{2}{3}\;\jbe_2(\ang)\,\mu^{\textrm{\tiny COS}}_0
- 4\, \jbe_1(\ang)\,  \mu^{\textrm{\tiny SIN}}_1\,, \\
\bar{g}_{\nud\ny} =&\, - \frac{2}{3}\;\jbe_2(\ang)\,\mu^{\textrm{\tiny SIN}}_0
+ 4\, \jbe_1(\ang) \, \mu^{\textrm{\tiny COS}}_1\,, \\
\bar{g}_{\nx\nx} =&\, \frac{1}{2} - \frac{1}{2} \jbe_0(\ang)\, \cos \left(\ang\right) 
- \jbe_1(\ang)\, \sin \left(\ang\right) \, \zeta^{\textrm{\tiny COS}}_2 \,,  \\
\bar{g}_{\nx\ny} =&\,  -\jbe_1(\ang)\, \sin \left(\ang\right) \, \zeta^{\textrm{\tiny SINCOS}}_1 \,, \\
\bar{g}_{\ny\ny} =&\, \frac{1}{2} - \frac{1}{2} \jbe_0(\ang)\, \cos \left(\ang\right)
- \jbe_1(\ang)\, \sin \left(\ang\right) \, \zeta^{\textrm{\tiny SIN}}_2 \,, 
\end{align}
\end{subequations}
using the following notations to simplify the expressions:
\begin{align}
  \mu_m &\equiv \frac{1}{N}\sum_{j=1}^{N} \left( \frac{t_j -t_0}{T} \right)^m \,, \\
  \mu^{\textrm{\tiny SIN}}_m &\equiv \frac{1}{N}\sum_{j=1}^{N} \left( \frac{t_j -t_0}{T} \right)^m \,
         \sin \left(\Omega\,t_j\right) \,, \\
  \mu^{\textrm{\tiny COS}}_m &\equiv \frac{1}{N}\sum_{j=1}^{N} \left( \frac{t_j -t_0}{T} \right)^m \,
        \cos \left(\Omega\,t_j\right) \,, \\
  \zeta^{\textrm{\tiny SIN}}_2    &\equiv \frac{1}{N}\sum_{j=1}^{N} \sin^2 \left(\Omega\,t_j \right) \, ,
  \;\;\; \zeta^{\textrm{\tiny COS}}_2   \equiv \frac{1}{N}\sum_{j=1}^{N} \cos^2 \left(\Omega\,t_j \right) \,,\\
   \zeta^{\textrm{\tiny SINCOS}}_1    &\equiv \frac{1}{N}\sum_{j=1}^{N} \sin \left(\Omega\,t_j \right)\,
   \cos \left(\Omega\,t_j \right) \,,
\end{align}
where $m$ can be zero or a positive integer number.

The components $\bar{g}_{ab}$ of the semicoherent metric tensor  are
also explicitly independent of the coordinates. Therefore,
the number of fine-grid templates $\Ntfine$ as described
by \Eref{e:Ntfine} is rewritten as
 \begin{align}
  \Ntfine &=  \rho_0\, \sqrt{\det \bar{g}} \int_{\Psearch}  \rmd \vDoppler \nonumber\\
  &= \Ntcoarse \sqrt{\frac{\det \bar{g}}{\det g^{[j]} }}\,.
  \label{e:Ntfine2}
\end{align}

Considering the distribution of the segments' midpoints~$\{t_j\}$, 
the quantities $\mu_m$ may be interpreted as the $m$th moment
of this distribution. Thus a very convenient choice of reference 
time~$t_0$ for the semicoherent metric
is given by the time average of all segment's midpoints~$\{t_j\}$,
\begin{equation}
   t_0=\frac{1}{N}\sum_{j=1}^{N}t_j \,.
   \label{e:t0center}
\end{equation}
With this choice of $t_0$ the quantities $\mu_m$ 
become the $m$th \emph{central} moments  (denoted by $\hat{\mu}_m$)
of the distribution of the segment midpoints~$\{t_j\}$.

In addition, for a large number of segments $N$, where the data set spans
over many cycles of the Earth's spinning motion with the one-day period $2\pi/\Omega$,
the following approximations may also be used:
\begin{subequations}
\label{e:LargeNapprox}
\begin{align}
  &\frac{1}{N}\sum_{j=1}^{N}  \sin \left(\Omega\,t_j\right) \approx 0\,,\qquad
  \frac{1}{N}\sum_{j=1}^{N}  \cos \left(\Omega\,t_j\right) \approx 0 \,, \\
  &\frac{1}{N}\sum_{j=1}^{N}  \sin^2 \left(\Omega\,t_j\right) \approx \frac{1}{2} \,, \qquad 
  \frac{1}{N}\sum_{j=1}^{N}  \cos^2 \left(\Omega\,t_j\right) \approx \frac{1}{2}\,,\\
  & \frac{1}{N}\sum_{j=1}^{N}  \sin \left(\Omega\,t_j\right) \,  \cos \left(\Omega\,t_j\right) \approx 0\,,
\end{align}
as well as
\begin{equation}
  \frac{1}{N}\sum_{j=1}^{N}  \frac{t_j}{T}\, \sin \left(\Omega\,t_j\right) \approx 0\,, \qquad
  \frac{1}{N}\sum_{j=1}^{N}  \frac{t_j}{T}\, \cos \left(\Omega\,t_j\right) \approx 0 \,.
\end{equation}  
\end{subequations}
Hence, given
\mbox{$\mu^{\textrm{\tiny SIN}}_0 \approx 0$},
\mbox{$\mu^{\textrm{\tiny COS}}_0 \approx 0$},
\mbox{$\mu^{\textrm{\tiny SIN}}_1 \approx 0$},
\mbox{$\mu^{\textrm{\tiny COS}}_1 \approx 0$},
\mbox{$\zeta^{\textrm{\tiny SIN}}_2 \approx 1/2$},
\mbox{$\zeta^{\textrm{\tiny COS}}_2 \approx 1/2$}, and
\mbox{$\zeta^{\textrm{\tiny SINCOS}}_1 \approx 0$},
the semicoherent metric components of \Esref{e:OneSdSemiCohMetricComps1}
along with the $t_0$ choice of \Eref{e:t0center} take the following \emph{diagonal} form:
\begin{equation}
\bar{g}  \approx  
\begin{pmatrix}
  \frac{1}{3} & 0 &  0 & 0 \smallskip\\
  0 &   \frac{4}{45}+ \frac{16}{3}\,\hat{\mu}_2  & 0 & 0 \smallskip\\
  0 & 0 & \frac{R(\ang)}{2} & 0\smallskip \\
  0 & 0 & 0 & \frac{R(\ang)}{2} \smallskip \\
\end{pmatrix} \,,
\label{e:OneSdSemiCohMetricComps2}
\end{equation}
where $R(\ang)$ is defined as
\begin{equation}
  R(\ang) \equiv 1 - \jbe_0(\ang)\, \cos \left(\ang\right) 
  - \jbe_1(\ang)\, \sin \left(\ang\right) \,.
  \label{e:Rofphi}
\end{equation}
The determinant of the above semicoherent metric tensor $\bar{g}_{ab}$ 
from \Eref{e:OneSdSemiCohMetricComps2} is obtained as
\begin{align}
   \det \bar{g} &\approx \left[ \frac{1}{135} + \frac{4}{9} \,\hat{\mu}_2 \right] R^2(\ang) \,.
    \label{e:detbargabSD1}
\end{align}
Finally, if $T$ is an integer multiple $q$ of one 
sidereal day,  \mbox{$T=\frac{2\pi}{\Omega}q$},  such that $\ang=\pi q$
and $R(\pi q)=1$,
the metric tensor~$\bar{g}$ from \Esref{e:OneSdSemiCohMetricComps2} 
simplifies to
\begin{equation}
\bar{g}  \approx 
\begin{pmatrix}
  \frac{1}{3} & 0 &  0 & 0 \smallskip\\
  0 &   \frac{4}{45}+ \frac{16}{3}\,\hat{\mu}_2  & 0 & 0 \smallskip\\
  0 & 0 & \frac{1}{2} & 0\smallskip \\
  0 & 0 & 0 & \frac{1}{2} \smallskip \\
\end{pmatrix} \,,
\label{e:OneSdSemiCohMetricComps3}
\end{equation}
and the corresponding determinant is simply given by
\begin{equation}
  \det \bar{g} \approx \frac{1}{135}  + \frac{4}{9} \,\hat{\mu}_2 \,.
  \label{e:detbargab}
\end{equation}

\subsection{Parameter-space resolution refinement\label{sssec:ParamRefinemOneSd}}

The results for the semicoherent metric tensor shown in 
\Esref{e:OneSdSemiCohMetricComps2} feature an important property:
$\bar{g}_{\nud\nud}$ represents the \emph{only} component of the 
semicoherent metric tensor $\bar{g}$ which 
significantly changes with an increased number of data segments $N$,
not converging to some constant value. 
In this respect, increasing~$N$ means that the number of fine-grid templates 
needs to be increased in only one dimension compared 
to a given coarse grid. 

To describe the refinement quantitatively, we use
\Eref{e:Ntfine2} to introduce the \emph{refinement factor}, 
denoted by $\gamma$, 
defining the ratio of the fine and coarse template-grid points:
\begin{equation}
  \gamma \equiv \frac{\Ntfine}{\Ntcoarse} = \sqrt{\frac{\det \bar{g} }{\det g^{[j]} }} 
   \,,
   \label{e:gamma-s1}
\end{equation}
where in the last step \Esref{e:Ntcoarse2} and \eref{e:Ntfine2} have been used.
In what follows, refinement factor \mbox{$\gamma_1$ ($\gamma_2$)} with 
the subscript~\mbox{$1$ ($2$)} is used to indicate the one-spindown 
(two-spindown) case, respectively.

Thus, by means of \Esref{e:detgabj0} and \eref{e:detbargabSD1} 
the refinement factor~$\gamma_1$ for the one-spindown case is explicitly given by
\begin{equation}
  \gamma_1 =   \sqrt{1 + 60\,\hat{\mu}_2} \; Q(\ang)\,,
  \label{e:gamma1}
\end{equation}
where  the function $Q(\ang)$ is defined as
\begin{align}
  Q(\ang) &\equiv R(\ang)\, \bigl[1 -6\,{\jbe_1}^2(\ang) - {\jbe_0}(\ang)\,\cos\left(\ang\right) \bigr]^{-1/2} \nonumber\\
 &\hspace{0.5cm}\times \bigl[ 1 - 10\,{\jbe_2}^2(\ang) - {\jbe_1}(\ang)\,\sin\left(\ang\right) \nonumber\\
 &\hspace{1cm}- {\jbe_0}(\ang)\,\cos\left(\ang\right) \bigr]^{-1/2} \,.
\end{align}
Note that from \Eref{e:gamma1} it is obvious that $\gamma_1$ 
scales linearly with the number of data segments~$N$, 
since only $\hat{\mu}_2$ depends (quadratically) on~$N$.
Hence, the enhanced parameter-space resolution 
resulting from the incoherent combination increases approximately
as $\propto N$ solely due to the first spindown parameter.
This is related to the fact that the number of possible (linear) spindown tracks
in frequency across all segments, of course, grows linearly with~$N$, too.

\subsection{Illustrative example\label{ssec:ExampleOneSd}}

\begin{figure}[t]
  \includegraphics[width=\columnwidth]{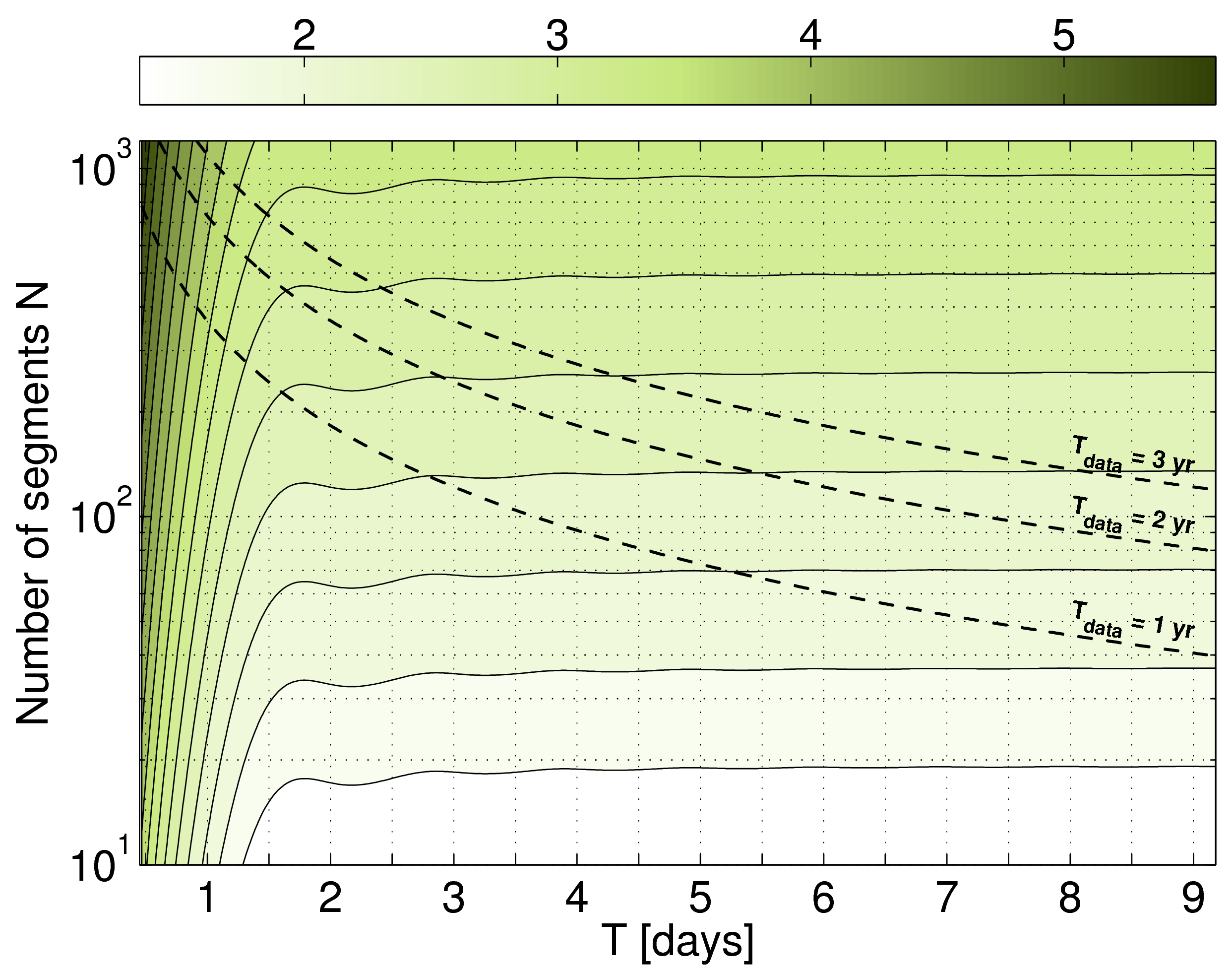}
  \caption{Refinement factor $\gamma_{1}$ for the
  one-spindown case. In this plot the color-coded contours
  show $\log_{10}(\gamma_{1})$ as a function of the coherent segment
  length $T$ and the number of segments~$N$.
  It has been assumed that there are no gaps between
  neighboring segments, and so \Eref{e:mu2simple} has been used. 
  The dashed lines mark locations of data sets with total time span 
  ($T_{\rm data}=NT$) of one, two, and three~years.
  \label{f:refinement-SD1} }
\end{figure}

As for a simple example, one may consider the case where $N$ segments
are uniformly distributed in such a way that there are no gaps between neighboring
segments, so that \mbox{$t_j=[j-(N+1)/2]\,T$}. 
It should be pointed out that this special case had been assumed \emph{a priori}
in the previous work of~\cite{bc2:2000}. 
Thus, for this particular instance the numerical  findings of~\cite{bc2:2000}
can be compared to the analytic results found here.
Denote the time span of the entire data set as \mbox{$T_{\rm data} \equiv N\, T$}.
Thus, in the present example the choice of \Eref{e:t0center} yields $t_0=0$ and
\begin{equation}
   \hat{\mu}_2 = \frac{N^2-1}{12} \,.
   \label{e:mu2simple}
\end{equation}
For this case,  $\gamma_{1}$ is shown in \Fref{f:refinement-SD1} 
for different values of $T$ and $N$.
For increasing values of~$T$, $\gamma_1$ 
rapidly converges to some upper-limit value for fixed $N$.
This maximum constant value can be arrived at 
analytically in the following way.
In this regime of $T$, $\det g^{[j]}$ from \Eref{e:detgabj0} can be
approximated by $1/135$ and $\det \bar{g}$ by \Eref{e:detbargab},
leading to $Q(\ang)\approx 1$ in \Eref{e:gamma1} and thus
\begin{equation}
  \gamma_1 \approx \sqrt{1 + 60\,\hat{\mu}_2}  =  \sqrt{5 N^2 - 4}  \,.
  \label{e:gamma1simpleA}
\end{equation}
where \Eref{e:mu2simple} has been used in the latter step.
This result agrees well with the corresponding refinement factor that can 
be computed from the numerical template counting formulas given in~\cite{bc2:2000}.

\section{Metric evaluation for  two spindown parameters \label{sec:TwoSpindowns}}

Searching for CW signals from potentially younger objects, 
spinning down faster, might eventually require one to also include a 
second spindown parameter (see \cite{Wette2008} for an example), 
taking also into account quadratic changes in frequency with time.
In this section, the additional semicoherent metric components 
are computed, which arise when a second spindown parameter 
is included in the search.
Thus, the coordinates $\{\nu,\nud,\nudd,\nx,\ny \}$ at time $t_0$ 
are used to label a point in this five-dimensional parameter space.
The phase model of \Eref{e:phase-new-coords}  with $s=2$ is given by
\begin{align}
   \Phi(t) &= \phi_0(t_0) 
   +  \nu(t_0) \,\frac{2\,(t-t_0)}{T}
   + \nud(t_0) \,\frac{4\,(t-t_0)^2}{T^2} \nonumber\\
   &+ \nudd(t_0) \,\frac{8\,(t-t_0)^3}{T^3} 
   + \nx(t_0)\,\cos \Omega\,t + \ny(t_0)\,\sin \Omega\,t \,.
   \label{e:phase-new-coords-2sd}
\end{align}
Based on \Eref{e:C1} the coordinates $\nu$, $\nud$, and $\nudd$ are 
explicitly written as
\begin{subequations}
\label{e:TwoSdNus}
\begin{align}
\nu(t_0) &= 2\pi\,\frac{T}{2} \biggl[ f(t_0) + f(t_0) \, \dot{\vec{\xi}}(t_0) \cdot \vec n \nonumber\\
  &\hspace{0.5cm} + \dot{f}(t_0)  \,\vec{\xi}(t_0) \cdot \vec n \biggr] \,, \\
\nud(t_0) &= 2\pi \left(\frac{T}{2}\right)^2
   \biggl[ \frac{\dot{f}(t_0)}{2} + \frac{f(t_0)}{2} \, \ddot{\vec{\xi}}(t_0) \cdot \vec n  \nonumber\\
  &\hspace{0.5cm}  + \dot{f}(t_0) \,  \dot{\vec{\xi}}(t_0) \cdot \vec n
   + \frac{ \ddot{f}}{2}\, \vec{\xi}(t_0) \cdot \vec n\biggr]  \,, \\
\nudd(t_0) &= 2\pi \left(\frac{T}{2}\right)^3
   \biggl[ \frac{\ddot{f}}{6} + \frac{f(t_0)}{6} \, \dddot{\vec{\xi}}(t_0) \cdot \vec n \nonumber\\
   &\hspace{0.5cm}    + \frac{\dot{f}(t_0)}{2} \, \ddot{\vec{\xi}}(t_0) \cdot \vec n
   + \frac{\ddot{f}}{2}\, \dot{\vec{\xi}}(t_0) \cdot \vec n  \biggr]  \,.
\end{align}
\end{subequations}
The coordinates $\nx$ and $\ny$ are as introduced in \Esref{e:skycoords}.

\subsection{Coherent metric\label{sssec:TwoSpindownCohMet1}}

Including a second spindown parameter $\nudd$ in the phase model
of \Eref{e:phase-new-coords} yields the following additional components for
 the coherent metric  tensor $g^{[j]}$: 
\begin{subequations}\label{e:TwoSdCohMetricComps1}
\begin{align}
g^{[j]}_{\nu\nudd} =&\, \frac{1}{5} + 4 \left( \frac{t_j-t_0}{T} \right)^2 \,, \\
g^{[j]}_{\nud\nudd} =&\, \frac{4}{3} \, \left( \frac{t_j-t_0}{T} \right) 
   +  16 \left( \frac{t_j-t_0}{T} \right)^3\,, \\
g^{[j]}_{\nudd\nudd} =&\, \frac{1}{7} + 8 \left( \frac{t_j-t_0}{T} \right)^2  
   + 48 \left( \frac{t_j-t_0}{T} \right)^4\,, \\
g^{[j]}_{\nudd\nx} =&\, \left[ - \frac{3}{5} \, \jbe_1(\ang) + \frac{2}{5} \, 
\jbe_3(\ang)\right] \sin\left(\Omega\,t_j\right) \nonumber\\
&\; - 4\,\jbe_2(\ang) \,\left(\frac{t_j - t_0}{T}\right)\,\cos\left(\Omega\,t_j\right) \nonumber\\
&\; - 12\,\jbe_1(\ang) \,\left(\frac{t_j - t_0}{T}\right)^2\,\sin\left(\Omega\,t_j\right) \,,\\
g^{[j]}_{\nudd\ny} =&\, \left[ \frac{3}{5} \, \jbe_1(\ang) - \frac{2}{5} \, 
\jbe_3(\ang)\right] \cos\left(\Omega\,t_j\right) \nonumber\\
&\; -4\,\jbe_2(\ang) \,\left(\frac{t_j - t_0}{T}\right)\,\sin\left(\Omega\,t_j\right) \nonumber\\
&\; + 12\,\jbe_1(\ang) \,\left(\frac{t_j - t_0}{T}\right)^2\,\cos\left(\Omega\,t_j\right) \,.
\end{align}
\end{subequations}
The components $g^{[j]}_{ab}$ of the coherent metric tensor  are
explicitly independent of the coordinates as mentioned earlier. 
Therefore,  the number of coarse-grid templates $\Ntcoarse$ 
is also computed as presented by \Eref{e:Ntcoarse2}.
To analytically estimate the actual value of $\Ntcoarse$ for the two-spindown
case, in addition to \Esref{e:range-nu} and \eref{e:range-nud} the 
following ranges of $\nudd$ are assumed to be searched
\begin{equation}
   -\pi\frac{T^3\,f}{12\,{\tau_{\tiny \textrm{min}}}^2}  \lesssim\; \nudd \;\lesssim
   \pi\frac{T^3\,f}{12\,{\tau_{\tiny \textrm{min}}}^2} \,.
   \label{e:range-nudd}
\end{equation}
Thus, evaluation of \Eref{e:Ntcoarse2} yields in this case
\begin{align}
 \Ntcoarse &=  \rho_0 \sqrt{\det g^{[j]}}
  \int_{\pi T f_{\tiny \textrm{min}}}^{\pi T f_{\tiny \textrm{max}}}  \rmd \nu 
  \int_{\mathcal{D}_f} \rmd\nx\, \rmd\ny \nonumber\\
  &\hspace{0.5cm}\times 
  \int_{-\frac{\pi T^2 f}{4\,\tau_{\tiny \textrm{min}}}}^{\frac{\pi T^2 f}{4\,\tau_{\tiny \textrm{min}}}} 
   \rmd \nud
   \int_{-\frac{\pi T^3 f}{12\,{\tau_{\tiny \textrm{min}}}^2}}^{\frac{\pi T^3 f}{12\,{\tau_{\tiny \textrm{min}}}^2}} 
   \rmd \nudd
   \nonumber\\
 &\approx  \rho_0 \sqrt{\det g^{[j]}}\,\frac{\pi^6 \tau_E^2}{15\,{\tau_{\tiny \textrm{min}}}^3} \,
  T^6 \left( f^5_{\tiny \textrm{max}} -f^5_{\tiny \textrm{min}}\right)\,. 
  \label{e:Ntcoarse1-2sd}
\end{align}

\begin{figure}[b]
  \includegraphics[width=0.9\columnwidth]{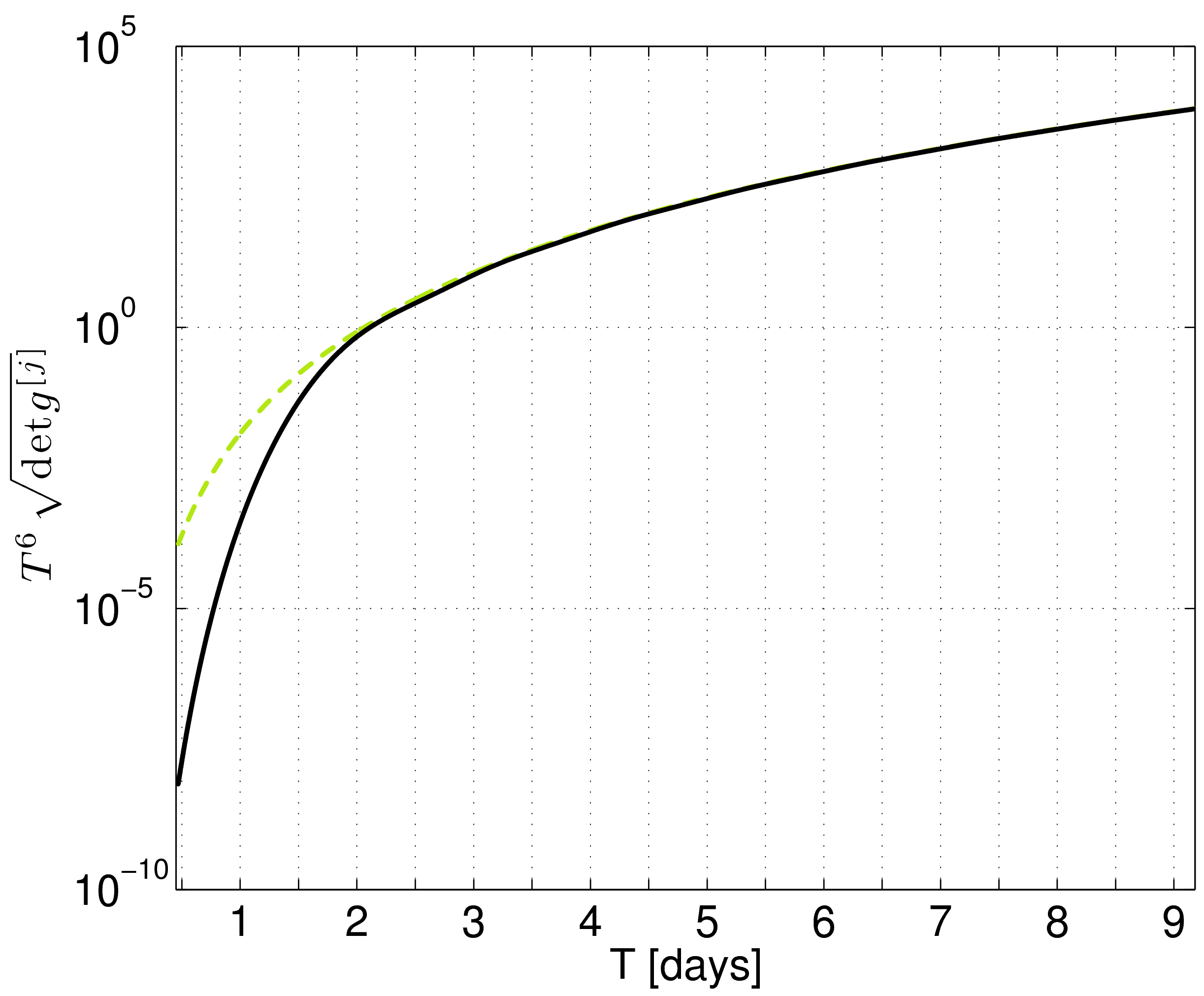}
  \caption{
  Dependency of the number of coarse-grid templates~$\Ntcoarse$
  on the coherent integration time $T$ (segment length) for the
  two-spindown case. 
  The solid curve shows $T^6\,\sqrt{\det g^{[j]}}$, since
  \mbox{$\Ntcoarse \propto T^6\,\sqrt{\det g^{[j]}}$}.
  The dashed curve shows~$T^6\sqrt{4/23625}$.
   \label{f:detgabj0-s1-SD2} }
\end{figure}

The determinant of the coherent metric tensor $g^{[j]}$ for
the two-spindown case is obtained accordingly as
\begin{align}
 \det\, &g^{[j]} =  \frac{4}{23625}\;
  \bigl[ 
  1+ 375\, {\jbe_1}^2(\ang)\, {\jbe_2}^2(\ang) + 189\,{\jbe_1}^4(\ang) \nonumber\\
  &- 252\,{\jbe_1}^3(\ang)\, \jbe_3(\ang) 
  + 84\,{\jbe_1}^2(\ang)\,{\jbe_3}^2(\ang) \nonumber\\
   &+ 42\,\jbe_1(\ang)\, \jbe_3(\ang)\
  - 42\,\jbe_0(\ang)\, \jbe_1(\ang)\, \jbe_3(\ang)\, \cos\!\left(\ang\right) \nonumber\\
  &+ 75 \, {\jbe_1}^3(\ang)\,\sin\!\left(\ang\right)/2 
  - 14\, {\jbe_3}^2(\ang) - 10\,{\jbe_2}^2(\ang) \nonumber\\
  &+ 14\,\jbe_0(\ang)\, {\jbe_3}^2(\ang)\, \cos\!\left(\ang\right)
   - 69\, {\jbe_1}^2(\ang)  \nonumber\\
  &+ 10\, \jbe_0(\ang)\, {\jbe_2}^2(\ang)\, \cos\!\left(\ang\right) 
  -2\,\jbe_0(\ang)\, \cos\!\left(\ang\right) \nonumber\\
  &+ 69\, \jbe_0(\ang)\, {\jbe_1}^2(\ang)\, \cos\!\left(\ang\right) 
  + {\jbe_0}^2(\ang)\, \cos^2\!\left(\ang\right) \nonumber\\
  & - \jbe_1(\ang)\,\sin\!\left(\ang\right)  
  + \jbe_0(\ang)\, \jbe_1(\ang)\, \sin\!\left(\ang\right)\,  \cos\!\left(\ang\right)
  \bigr] \,.
  \label{e:detgabj0-SD2}
\end{align}
As mentioned earlier, $\det g^{[j]}$ only depends upon $\ang$.
To investigate the scaling of $\Ntcoarse$ with the coherent integration time~$T$,
\Fref{f:detgabj0-s1-SD2} shows $T^6\sqrt{\det g^{[j]}}$,
being a measure for~$\Ntcoarse$, versus~$T$. 
For increasing values of~$T$, $\det g^{[j]}$ 
converges to the constant value of~$4/23625$,
since the metric tensor components related to the Earth's spinning
motion, $\nx$ and $\ny$,
become approximately constant in this regime as explained
earlier in \Sref{ssec:OneSpindownCohMet}.

Finally, it should be pointed out that choosing \mbox{$t_0=t_j=0$} 
simplifies the expressions of \Esref{e:TwoSdCohMetricComps1},
yielding
\begin{subequations}\label{e:TwoSdCohMetricComps2}
\begin{align}
g^{[j=0]}_{\nu\nudd} =&\, \frac{1}{5} \,, \\
g^{[j=0]}_{\nud\nudd} =&\, 0 \,, \\
g^{[j=0]}_{\nudd\nudd} =&\, \frac{1}{7} \,, \\
g^{[j=0]}_{\nudd\nx} =&\, 0 \,,\\
g^{[j=0]}_{\nudd\ny} =&\, \left[ \frac{3}{5} \, \jbe_1(\ang) - \frac{2}{5} \, 
\jbe_3(\ang)\right]  \,.
\end{align}
\end{subequations}

\subsection{Semicoherent metric\label{sssec:TwoSpindownSemiCohMet1}}

The extra components of the semicoherent metric tensor $\bar{g}$ are
then obtained via \Eref{e:avegab} using \Esref{e:TwoSdCohMetricComps1} as
\begin{subequations}\label{e:TwoSdSemiCohMetricComps1}
\begin{align}
\bar{g}_{\nu\nudd} =&\, \frac{1}{5}+ 4\,\mu_2  \,, \\
\bar{g}_{\nud\nudd} =&\,\frac{4}{3}\,\mu_1+ 16\,\mu_3 \,, \\
\bar{g}_{\nudd\nudd} =&\, \frac{1}{7} + 8\,\mu_2 + 48\,\mu_4 \,, 
\end{align}
\begin{align}
\bar{g}_{\nudd\nx} =&\, \left[- \frac{3}{5} \, \jbe_1(\ang) + \frac{2}{5} \, \jbe_3(\ang)\right] 
\,\mu^{\textrm{\tiny SIN}}_0 \nonumber\\
&\;- 4\,  \jbe_2(\ang)\,  \mu^{\textrm{\tiny COS}}_1
- 12\,  \jbe_1(\ang)\,  \mu^{\textrm{\tiny SIN}}_2 \,, \\
\bar{g}_{\nudd\ny} =&\, 
\left[ \frac{3}{5} \, \jbe_1(\ang) - \frac{2}{5} \, \jbe_3(\ang)\right]
\,\mu^{\textrm{\tiny COS}}_0 \nonumber\\
&\;- 4\,  \jbe_2(\ang)\,  \mu^{\textrm{\tiny SIN}}_1
+ 12\,  \jbe_1(\ang) \,  \mu^{\textrm{\tiny COS}}_2 \,.
\end{align}
\end{subequations}
With the approximations of \Esref{e:LargeNapprox} used earlier, 
\mbox{$\mu^{\textrm{\tiny SIN}}_0 \approx 0$},
\mbox{$\mu^{\textrm{\tiny COS}}_0 \approx 0$},
\mbox{$\mu^{\textrm{\tiny SIN}}_1 \approx 0$},
\mbox{$\mu^{\textrm{\tiny COS}}_1 \approx 0$},
and the $t_0$ choice of \Eref{e:t0center}
the semicoherent metric tensor components in \Esref{e:TwoSdSemiCohMetricComps1} 
take the following form
\begin{subequations}\label{e:TwoSdSemiCohMetricComps2}
\begin{align}
\bar{g}_{\nu\nudd} =&\, \frac{1}{5}+ 4\, \hat{\mu}_2  \,, \\
\bar{g}_{\nud\nudd} \approx&\, 0 \,, \\
\bar{g}_{\nudd\nudd} =&\, \frac{1}{7} + 8\, \hat{\mu}_2 + 48\, \hat{\mu}_4 \,, \\
\bar{g}_{\nudd\nx} \approx &\, -6\, \jbe_1(\ang) \,  \hat{\mu}^{\textrm{\tiny COS}}_2 
\approx  -6\,  \jbe_1(\ang) \,  \hat{\mu}_2 \,, \\
\bar{g}_{\nudd\ny} \approx&\, 6\,  \jbe_1(\ang)\,  \hat{\mu}^{\textrm{\tiny SIN}}_2
\approx  6\,  \jbe_1(\ang)\, \hat{\mu}_2 \,.
\end{align}
\end{subequations}
Thus, the full five-dimensional semicoherent metric tensor $\bar{g}$ is 
obtained as
\begin{widetext}
\begin{equation}
\bar{g}  \approx 
\begin{pmatrix}
  \frac{1}{3} & 0 &  \frac{1}{5}+ 4\,\hat{\mu}_2 & 0 & 0 \smallskip\\
  0 &   \frac{4}{45}+ \frac{16}{3}\,\hat{\mu}_2  & 0 & 0 & 0 \smallskip\\
  \frac{1}{5}+ 4\,\hat{\mu}_2 & 0 &  \frac{1}{7} + 8\,\hat{\mu}_2 + 48\,\hat{\mu}_4 & -6\,  \jbe_1(\ang) \,  \hat{\mu}_2  & 6\,  \jbe_1(\ang) \,  \hat{\mu}_2 \smallskip \\
  0 & 0 &  -6\,  \jbe_1(\ang) \,  \hat{\mu}_2 &  \frac{R(\ang)}{2} & 0\smallskip \\
  0 & 0 & 6\, \jbe_1(\ang) \,  \hat{\mu}_2 & 0 & \frac{R(\ang)}{2} \smallskip \\
\end{pmatrix} \,,
\label{e:TwoSdSemiCohMetricComps4}
\end{equation}
\end{widetext}
where $R(\ang)$ is given by \Eref{e:Rofphi}.
The corresponding determinant is obtained as
\begin{align}
 \det \bar{g} &\approx
 \frac{1+60\, \hat{\mu}_2}{23625}\,4R^2(\ang)\,  \nonumber\\ 
 &\;\;\; \times140\bigl[ 15 (\hat{\mu}_4 - {\hat{\mu}_2}^2) 
 -45\frac{ {\jbe_1}^2(\ang)\, {\hat{\mu}_2}^2}{R(\ang)}
 + \hat{\mu}_2 \bigr] \,.
 \label{e:detbargab-SD2}
\end{align}

When $T$ is an integer multiple $q$ of one 
sidereal day,  \mbox{$T=\frac{2\pi}{\Omega}q$},  
such that $\ang=\pi q$ and $R(\pi q)=1$,
the determinant of the semicoherent metric tensor takes the form
\begin{align}
 \det \bar{g} &\approx 
 \frac{1+60\, \hat{\mu}_2}{675}  16\bigl[ 15 (\hat{\mu}_4 - {\hat{\mu}_2}^2) 
 -\frac{45\, {\hat{\mu}_2}^2}{\pi^2 q^2}
 + \hat{\mu}_2 \bigr] \,.
  \label{e:detbargab-SD2-q}
\end{align}

\subsection{Parameter-space resolution refinement\label{sssec:ParamRefinemTwoSd}}

The refinement factor for the two-spindown parameter case~$\gamma_2$ 
has been defined through \Eref{e:gamma-s1} as
\begin{equation}
  \gamma_2 = \sqrt{\frac{\det \bar{g} }{\det g^{[j]} }} \,.
\end{equation}
Here $g^{[j]}$ and $\bar{g}$ denote the coherent and
semicoherent metric tensors, respectively, for the two-spindown parameter case.
Substituting $\det g^{[j]}$ by \Eref{e:detgabj0-SD2}  and $\det \bar{g}$ by \Eref{e:detbargab-SD2} yields
\begin{align}
  \gamma_2 &= 2\sqrt{35}\,\sqrt{1+60\, \hat{\mu}_2} \,  \nonumber\\ 
 &\;\;\; \times\left[ 15 (\hat{\mu}_4 - {\hat{\mu}_2}^2) 
 -45\frac{ {\jbe_1}^2(\ang)\, {\hat{\mu}_2}^2}{R(\ang)}
 + \hat{\mu}_2 \right]^{1/2} \; U(\ang) \,,
 \label{e:gamma2}
\end{align}
where the explicit expression for~$U(\ang)$ can be 
deduced from \Eref{e:detgabj0-SD2} (suppressed
for brevity here). 

From \Eref{e:gamma2} the scaling of $\gamma_2$ at leading order in~$N$ is
obtained as  \mbox{$\gamma_2 \propto N^3$}, using \mbox{$\hat{\mu}_4 \propto N^4$}
and \mbox{$\hat{\mu}_2 \propto N^2$}. This cubic scaling with~$N$ 
is solely due to the first and second spindown parameters,
resulting from the product of possible linear and quadratic spindown 
tracks in frequency across the segments, which obviously grows as $N^3$.

Note that in general, for $s$ spindown parameters, the expected scaling of the 
refinement factor  $\gamma_{s}$ with the number of data segments~$N$ is 
given by 
\begin{equation}
  \gamma_{s} \propto N^{s(s+1)/2} \,.
\end{equation}
This scaling with~$N$ is robust and agrees with what can be deduced 
from the numerical findings of Ref.~\cite{bc2:2000}.

\subsection{Illustrative example\label{ssec:ExampleTwoSd}}

To further examine~$\gamma_2$,
it is instructive to consider again the example data set presented 
in \Sref{ssec:ExampleOneSd}.
Thus, given \mbox{$t_j=[j-(N+1)/2]\,T$} and $t_0=0$, the 
third central moment vanishes, $\hat{\mu}_3=0$,
and the fourth central moment $\hat{\mu}_4$ is obtained as
\begin{equation}
   \hat{\mu}_4 =\frac{N^4}{80} - \frac{N^2}{24} + \frac{7}{240} \,.
   \label{e:mu4simple}
\end{equation}
Furthermore, it holds
\begin{equation}
   \hat{\mu}_4 - {\hat{\mu}_2}^2 =\frac{N^4}{180} - \frac{N^2}{36} + \frac{1}{45} \,.
   \label{e:mu4-mu22}
\end{equation}

\begin{figure}[b]
  \includegraphics[width=\columnwidth]{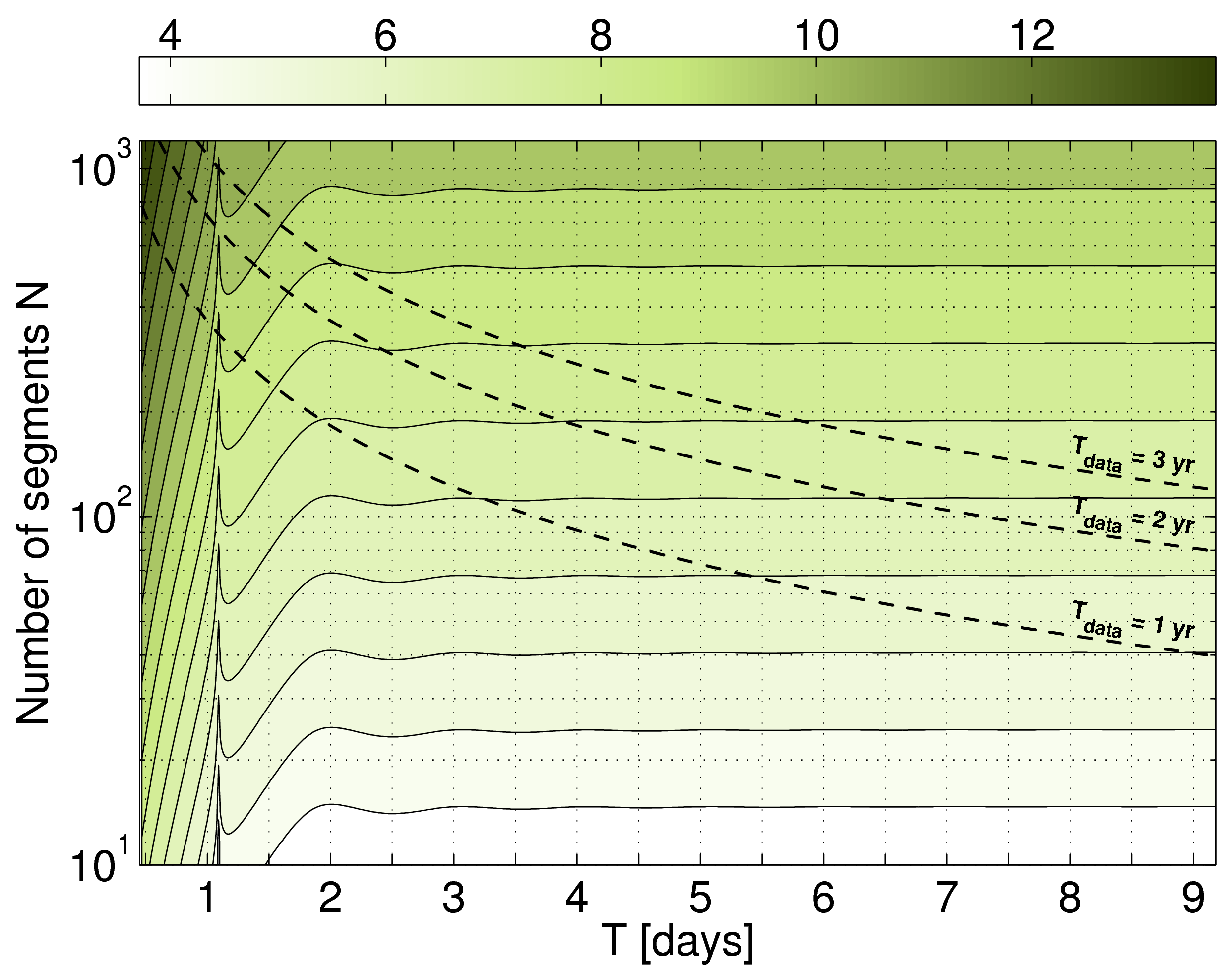}
  \caption{Refinement factor $\gamma_{2}$ for the
  two-spindown case. In this plot the color-coded contours
  show $\log_{10}(\gamma_{2})$ as a function of the coherent segment
  length $T$ and the number of segments~$N$.
  It has been assumed that there are no gaps between
  neighboring segments. 
  The dashed lines mark locations of data sets with total time span 
  ($T_{\rm data}=NT$) of one, two, and three~years.
  \label{f:refinement-SD2} }
\end{figure}

Using  \Eref{e:mu4simple} along with \Eref{e:mu2simple} the
refinement factor~$\gamma_2$ can be computed as a  function
of~$T$ and~$N$, as illustrated in \Fref{f:refinement-SD2}.
For increasing values of $T$, $\gamma_2$ 
rapidly converges to some constant value for fixed $N$.
In this case, $\det g^{[j]}$ of \Eref{e:detgabj0-SD2} is well
approximated by $4/23625 $ and $\det \bar{g}$ of \Eref{e:detbargab-SD2-q} by
\begin{equation}
   \det \bar{g} \approx 
   \frac{240}{675}\, ( 1+60\,\hat{\mu}_2)\,
    (\hat{\mu}_4 - {\hat{\mu}_2}^2)  \,.
\end{equation}
Hence, $\gamma_2$ in this case is described by
\begin{align}
  \gamma_2 &\approx \sqrt{2100\,(1 + 60\,\hat{\mu}_2)\, (\hat{\mu}_4 - {\hat{\mu}_2}^2)} \nonumber \\
   &\approx \gamma_1\, \sqrt{(35\,N^4 - 175\,N^2 + 140)/3} \,.
  \label{e:gamma2simple}
\end{align}
where \Esref{e:mu2simple} and (\ref{e:mu4-mu22}) have been used.
Thus, the anticipated scaling at leading 
order in~$N$ is recovered: $\gamma_2 \propto N^3$.

\section{Conclusion\label{sec:Conclusion}}

A formalism has been presented for the parameter-space metric
of semicoherent CW searches, where the data are divided into
segments that are coherently analyzed and subsequently
combined incoherently. 
By using new coordinates on parameter space,
the first fully analytical semicoherent metric for broadband all-sky 
CW surveys has been derived. 
Additionally, in the new coordinates the components of both 
the coherent and the semicoherent metric tensor are constant, 
being explicitly independent of the coordinates. This entails
great convenience regarding practical aspects
of semicoherent CW searches.

Explicit analytic expressions of the semicoherent
metric tensor components have been obtained for
two typical search parameter spaces of current practical interest.
First, the one-spindown case has been considered,
restricting to linear frequency drifts with time as done in
many current all-sky searches. Second, the
semicoherent metric also has been calculated and examined for
the two-spindown case, where additionally 
quadratic changes in frequency are taken into account.

Analytic signal template counting formulas have
been provided for the coherent stage (coarse grid)
as well as for the incoherent combination step (fine grid). 
In this respect, a useful quantity, called the \emph{refinement factor},
has been introduced as the ratio of the number of fine-grid 
and coarse-grid templates. Thus, the refinement factor describes
(coordinate independently)
the additional parameter-space metric resolution gained
from the combination of segments. Moreover, 
the scaling of the refinement factor with the number of segments 
has been found to be predominantly determined 
solely by the spindown parameters. 

The present results also embed the case of directed 
semicoherent searches,
where the sky position is known and hence is not a search
parameter. Thereby, the search parameter space consists
only of frequency and spindown parameters. 
The corresponding semicoherent metric tensor
components are identical to ones derived here.
The resulting scalings of the refinement factor with 
the number of segments also hold, since governed by 
the spindown parameters as described above.

The formalism presented in this paper assumes
segments that are very short compared to one year.
The Earth's orbital motion then can be described
by a low-order Taylor expansion, and therefore be modeled
by changes in frequency and frequency derivatives.
In the future, however, increased computing power might
allow one to use segments  substantially longer than a few days 
and hence issues remain to be explored in such circumstances.

\newpage

\section{Acknowledgments}
I am indebted to Bruce Allen, Reinhard Prix, and Chris Messenger for 
very useful discussions and helpful comments on this work. 
I also thank Miroslav Shaltev for corrections of the manuscript.
I gratefully acknowledge the support of the Max-Planck-Society.
This document has been assigned LIGO Document Number~\dcc.

\bibliography{SemiCohMetr}

\end{document}